\documentclass[12pt,preprint]{aastex}

\shortauthors{C.W. Lee et al.}
\shorttitle{INTERNAL MOTIONS IN STARLESS DENSE CORES}

\slugcomment{Accepted for publication in The Astrophysical Journal}
\newcommand{\skipthis}[1]{}
 \def\today{\rightline{\ifcase\month\or
         January\or February\or March\or April\or May\or June\or
         July\or August\or September\or October\or November\or December\fi
         \space\number\day, \number\year}}
\def\etal{{\it et al.}\ }

\def\msun{{\rm\,M_\odot}}

\def\s-1{{\rm\,s^{-1}}}

\def\spose#1{\hbox to 0pt{#1\hss}}
\def\C3H2{{\rm\,\rm C_3H_2}}
\def\NH3{{\rm\,\rm NH_3}}
\def\HOCO+{{\rm\,\rm HOCO^+}}
\def\lta{\mathrel{\spose{\lower 3pt\hbox{$\mathchar"218$}}
     \raise 2.0pt\hbox{$\mathchar"13C$}}}
\def\gta{\mathrel{\spose{\lower 3pt\hbox{$\mathchar"218$}}
     \raise 2.0pt\hbox{$\mathchar"13E$}}}

\begin{document}

\font\twelvei = cmmi10 scaled\magstep1 
       \font\teni = cmmi10 \font\seveni = cmmi7
\font\mbf = cmmib10 scaled\magstep1
       \font\mbfs = cmmib10 \font\mbfss = cmmib10 scaled 833
\font\msybf = cmbsy10 scaled\magstep1
       \font\msybfs = cmbsy10 \font\msybfss = cmbsy10 scaled 833
\textfont1 = \twelvei
       \scriptfont1 = \twelvei \scriptscriptfont1 = \teni
       \def\mit{\fam1 }
\textfont9 = \mbf
       \scriptfont9 = \mbfs \scriptscriptfont9 = \mbfss
       \def\bmit{\fam9 }
\textfont10 = \msybf
       \scriptfont10 = \msybfs \scriptscriptfont10 = \msybfss
       \def\bmsy{\fam10 }

\def\etal{{\it et al.~}}
\def\eg{{\it e.g.}}
\def\ie{{\it i.e.}}
\def\lsim{\raise0.3ex\hbox{$<$}\kern-0.75em{\lower0.65ex\hbox{$\sim$}}} 
\def\gsim{\raise0.3ex\hbox{$>$}\kern-0.75em{\lower0.65ex\hbox{$\sim$}}} 
\title{Internal Motions in Starless Dense Cores}

\author{Chang Won Lee$^{1}$ and  Philip C. Myers$^{2}$}

\vskip 0.2in
\affil{$^1$ Korea Astronomy \& Space Science Institute,
776 Daedeokdae-ro, Yuseong-gu, Daejeon 305-348, Republic of Korea
E-mail: cwl@kasi.re.kr}

\vskip 0.2in
\affil{$^2$Harvard-Smithsonian Center for Astrophysics,
60 Garden Street, Cambridge, MA  02138, USA}

\vskip 1in
\begin{abstract}
This paper discusses the statistics of internal motions in starless dense cores and the relation of these motions to core density and evolution.
Four spectral lines from three molecular species are analyzed from single-pointing and mapped observations of several tens of starless cores.
Blue asymmetric profiles are dominant, indicating that inward motions are prevalent in sufficiently dense starless cores.
These blue profiles are found to be more abundant, and their asymmetry is bluer, 
at core positions with stronger $\rm N_2H^+$ line emission or higher column density.
Thirty three starless cores are classified into four types according to the blueshift and red shift of the lines
in their molecular line maps.
Among these cores, contracting motions dominate: 19 are classified as contracting, 3 as oscillating, 3 as expanding, and 8 as static.
Contracting cores have inward motions all over the core with predominance of those motions
near the region of peak density.
Cores with the bluest asymmetry tend to have greater column density than other cores and all five cores with peak column density
$> \rm 6\times 10^{21}~cm^{-2}$  are found to be contracting. This suggests that starless cores are likely to have contracting
motions if they are sufficiently condensed. 
Our classification of the starless cores may indicate a sequence of core evolution in the sense that column density
increases from static to contracting cores:
the static cores in the earliest stage, the expanding and/or the oscillating cores in the next, and the contracting cores in the latest stage.


\end{abstract}


\clearpage

\section{Introduction}

``Starless'' dense  ($\rm n_{H_2} \ge 10^4 cm^{-2}$) cores are the cores which have no embedded protostars or no associated T-Tauri stars
(Lee \& Myers 1999).
Such cores have no internal outflows and are well separated from external triggering effects such as outflows and winds from other stars, 
or supernova explosions. Therefore they are the best laboratory to explore initial conditions of isolated star formation
(e.g., Ward-Thompson 2002).
Internal motions in starless cores can indicate contraction, a sign of physical progress toward star formation.  
Ambipolar diffusion or dissipation of turbulence may drive
such motions (e.g.,Ciolek \& Mouschovias 1995; Nakano 1998; Myers \& Lazarian 1998). 
However, it is not conclusive yet which mechanisms bring the collapse of the cores to 
eventually form a protostar in the cores.

The presence of contracting or expanding motions along the line of sight has been inferred from spectroscopic observations 
of optically thick and thin molecular lines.
The ``spectral infall asymmetry'' is a combination of a pair of spectral line shapes. 
A double-peaked shape with a blue peak brighter than a red peak occurs in a thick line, while a single Gaussian-like shape occurs in a thin line. 
This combination is often used to trace ``inward'' motions of gaseous material in starless cores (Leung \& Brown 1977; Zhou 1995; Myers et al. 1996). 
Since the infall asymmetry in starless core was first reported in L1544 (Tafalla et al. 1998), it has been
observed in numerous starless dense cores  using
thin (such as $\rm N_2H^+$ 1-0) and thick (such as CS 2-1, CS 3-2, and HCN 1-0) molecular lines,
finding that inward motions are a dominant feature in starless cores
(Lee et al. 1999, 2004, Sohn et al. 2007).

Mapping observations of starless cores also indicated that the infall asymmetry is spatially extended comparable to the size of
the $\rm N_2H^+ $ cores (Tafalla et al. 1998; Lee et al. 2001; hereafter LMT01).
The most extensive mapping survey of starless cores to study a pattern of inward motions was given
by LMT01. Out of 53 starless cores observed 
in both of $\rm N_2H^+$ 1-0 and CS 2-1 lines, 19 infall candidates were selected
according to the spectral shapes of CS 2-1 lines, and the velocity shift of their brighter component from
the velocity of the systemic component traced by the optically thin  $\rm N_2H^+ $ line.
This survey showed that extended inward motions are a frequently occurring feature in starless cores, and
are probably a necessary process in the condensation of a star-forming dense core.

On the other hand,  mapping studies have also indicated that gaseous motions could be more complex in some cores than ``all inward''.
Some starless cores showed overabundance of red asymmetric profiles 
(e.g., L429-1 and CB246 from LMT01 and FeSt1-457 from Aguti et al. 2007). 
Just as the blue asymmetric profiles indicate inward motions, the cores showing red asymmetric profiles are likely to be expanding. 
The origin of such expansion is unclear. 
One suggested idea for the existence of red asymmetric profiles in a dense core is that 
some disturbance in the external pressure of the core under the state of
gravitational equilibrium (such as shock waves caused by nearby OB stars or supernovae) may trigger
its oscillatory motions in the outer layers and the core is now being observed at the status of expanding motion
(e.g., Lada et al. 2003, Redman et al. 2006, Aguti et al. 2007).
Other starless cores (L1495A-N, L1507A, and L1512 from LMT01 and  B68 from Redman et al. 2006) show a complicated mixture 
of blue and red asymmetric profiles. Oscillating motions of some specific mode in gaseous outer layers of the cores were also 
suggested to explain this feature (e.g., Lada et al. 2003, Broderick \& Keto 2010). 

Of course there are some cores which do not show any significant asymmetric feature in lines, i.e., no 
evidence for large scale contraction or expansion. 

The foregoing results raise several basic questions about core motions and about how such motions are distributed in a sample of cores.
How  likely is it for a core to have motions dominated by  
infall, expansion, oscillation, or no obvious pattern ? 
Do the patterns of the internal motions in starless cores reflect any evolutionary status of the cores toward the star formation ? 
Would some environmental conditions affect their internal motions in starless cores ?
Which physical parameters are significant, and how do they relate to observed internal motions ?
Answers for these questions may help us to improve our understanding of how stars form in dense cores.

For this study we compiled and analyzed molecular line data available to assess the statistics of internal motions of 
starless dense cores and discuss on how they evolve toward star formation.
In the following section the data that we collect are summarized. In Section 3 we examine 
the distribution of the normalized velocity differences $\rm \delta V$ between the optically thick and thin lines and its relation with physical quantities.
In the last section we discuss implication of the distribution of the $\rm \delta V$ in the context of dense core evolution
and environmental effect on the  $\rm \delta V$ distribution.  

\section{Data}
The spectral line data that are used in this study have been obtained from several previous observations, 
single pointing surveys of starless cores in CS 2-1 and $\rm N_2H^+$ 1-0 (Lee et al. 1999),  
in CS 3-2 (Lee et al. 2004), in HCN 1-0 (Sohn et al. 2007), and  a mapping survey in CS 2-1
and $\rm N_2H^+$ 1-0 (LMT01).
For these systematic surveys of starless cores we first constructed a catalog of optically selected cores
(Lee \& Myers 1999) from which 306 cores were selected as ``starless'' in the sense that they do not have either an embedded 
IRAS point source or a pre-main-sequence star.\footnote[1]
{Recent $\it Spitzer$'s Legacy Project ``From Molecular Cores to Planet Forming Disks'' (c2d; Evans et al.
2003) for small isolated cores indicated about 20$\%$ of ``starless'' cores harbor a very faint $\it Spitzer$ source which is called as
Very Low Luminosity Objects (VeLLOs;  Dunham et al. 2008).  We dropped such cores in any number statistics in this paper.}  
 
Next we performed systematic spectral line surveys of 220 starless cores listed in the catalog in  CS 2-1 and/or $\rm N_2H^+$ 1-0 
to sample 66 cores detected in both lines  (Lee et al. 1999).
Then we made observations of the cores in CS 3-2 to have 
66 cores detected in both CS 3-2 and $\rm N_2H^+$ (Lee et al. 2004), and also in HCN 1-0 to have 48 cores detected in both HCN and $\rm N_2H^+$
(Sohn et al. 2007).
Mostly the same sources were detected in all four set of molecular lines CS 2-1, 3-2, HCN 1-0 and $\rm N_2H^+$ 1-0 except for a few sources.
For example the number of cores detected in whole set of CS 2-1, 3-2 and $\rm N_2H^+$ 1-0 is 64. Thus two sources were detected 
in either CS 2-1 and $\rm N_2H^+$ 1-0 only or  CS 3-2 and $\rm N_2H^+$ 1-0 only . Out of 48 sources detected in HCN 1-0 and $\rm N_2H^+$ 46 cores were 
also detected in the set of CS and $\rm N_2H^+$ lines. This indicates that starless cores detected in CS, HCN, and $\rm N_2H^+$ 
that are discussed for various statistics from different molecular line observations in this paper are mostly the same.  

The mapping survey of a total of 53 targets has been performed with CS 2-1 and/or $\rm N_2H^+$ 1-0 to have 34 starless cores mapped 
in both lines (LMT01).
These data are used in the analysis of the spectral asymmetry.
The molecular line observations we collect for this study are summarized in Table 1.
  
The cores studied here have properties of typical mean density of $\rm \sim 10^4~cm^{-3}$ and column density of a few $\rm 10^{21}~cm^{-2}$
(Lee \& Myers 1999).
The mapped radii of the cores in $\rm N_2H^+$ lines vary from 0.1 to 0.3 pc with a typical value of $\sim 0.1$ pc.
The cores traced in CS lines are more extended, sometimes twice more, than $\rm N_2H^+$ cores. 
In Table 2 we collect detailed information on the starless cores that are frequently referred here, mostly from LMT01 regarding 
the degree of asymmetry in profiles and physical properties of the cores.
 
\section{Analysis of the Normalized Velocity Differences $\rm \delta V$ of Spectra}

Quantifying properly the spectral asymmetry is very important to infer which internal motions are dominant in starless cores.
The amount by which the optically thick spectrum is blueshifted or redshifted with respect to the 
optically thin line (N$_2$H$^+$ 1-0) can be estimated by using the normalized velocity difference 
$ \delta V = (V_{thick}-V_{thin})/\Delta V_{thin}$, where
$V_{thick}$ and $V_{thin}$ are the peak velocities of the optically thick and thin lines,
and $\Delta V_{thin}$ is the FWHM of the thin line (Mardones et al. 1997, Lee et al. 1999, LMT01).
All the velocity information of the spectra was obtained with a Gaussian fit of the lines. This $\rm \delta V$ is a useful quantity which can help 
to determine the degree of spectral asymmetry. For example, if the $ \delta V$ for the spectra is negative, 
this would indicate that the thick profile is in infall (or blue) asymmetry possibly tracing inward gaseous motions, called as a blue profile.
On the other hand, if the $ \delta V$ is positive, this may mean that the profile is now in outflow (or red) asymmetry tracing outward motions, 
called as a red profile.   

\subsection{Number Distribution of $\delta V$ of Optically Thin and Thick Spectra}
The number distribution of $\delta V$ of the observed spectra has been examined with single pointing observations toward central regions of the cores
using molecular lines of CS 2-1 (Lee et al. 1999),  CS 3-2 (Lee et al. 2004), 
and  HCN 1-0 (Sohn et al. 2007) as optically thick tracers and $\rm N_2H^+$ 1-0
as an optically thin tracer.
From the single-pointing survey for 220 starless cores in the lines of CS 2-1 and/or $\rm N_2H^+$ 1-0,
the  $\delta V$ distribution has been constructed for 66 cores detected in both CS 2-1 and $\rm N_2H^+$ 1-0 lines (Lee et al. 1999).
The  $\delta V$ distribution from single pointing observations for CS 3-2 and $\rm N_2H^+$ 1-0 
has also been tested for 66 starless cores (Lee et al. 2004; hereafter LMP04). 
Both $\delta V$ distributions for CS 2-1 and CS 3-2 with respect to the  $\rm N_2H^+$ 1-0
are found to be significantly skewed to the negative value, suggesting that CS 2-1 and  3-2 lines preferentially trace inward motions 
in the cores (see Figure 3 of LMP04 for the revised $\delta V$ distributions). 
The  $\delta V$ distribution for HCN 1-0 using its single pointing observations 
showed that the degree of the skewness in the distribution to the blue (the negative value) is even greater than that for any
other infall tracers such as CS 2-1 and  3-2, indicating that HCN is the best tracer of inward motions (Sohn et al. 2007).   
The skewed distribution of $\delta V$ to the blue for all tracers 
from previous single dish surveys implies that it is very likely that the central region of a core has 
inward gaseous motions.

Then how is it likely for a core to show infall asymmetry at any place when one looks at the dense core ? 
What kind of motions are more likely to be observed in a dense core ? 
The $\delta V$ analysis using all mapping observations with the above infall tracers may be able to address these questions.
For this purpose we use CS 2-1 and  $\rm N_2H^+$ 1-0 mapping data for 34 starless cores by LMT01 which 
give the most extensive data set of the line spectra available for this $\delta V$ analysis. 

Figure 1(a) is a histogram for the number distribution of  $\delta V$ for all observed positions of 34 starless  $\rm N_2H^+$ cores, indicating 
overabundance of blue profiles.
Note that half of the sample (17 cores; the group A) has been selected as infall candidates based
on the single-pointing observations of Lee et al. (1999) while the other half of the sample (the group B) has been 
selected because they were either bright in high density tracers such as CS,  $\rm N_2H^+$, and  $\rm NH_3$ lines, or 
very opaque in optical (LMT01). 
Figure 1(b) is the number distribution of  $\delta V$ for the cores for the group A
and Figure 1(c) is that for the group B. 
These figures are intended to test whether the $\delta V$ distribution of all cores shown in Figure 1(a) is significantly affected by a bias of sample selection.
The mean value (and standard error of the mean, hereafter s.e.m.) of the  $\delta V$ distribution for the entire sample is -0.2$\pm$0.02.
The mean value of $\delta V$ for the group of infall candidates of Figure 1(b) is  -0.26$\pm$0.02 
while that for no infall-biased sample of Figure 2(c) is  -0.12$\pm$0.03. 
This suggests that the  $\delta V$ distribution of the entire sample is somewhat affected by that of the the infall candidates group A.
Nonetheless, it should be noted that the $\delta V$  distribution of the group B also exhibits a statistically significant overabundance of blue profiles.

In fact,  the selection of our sample has been originally made from single pointing observations toward 220 starless cores 
with high density tracers such as CS and $\rm N_2H^+$ among which 66 cores were both detected with two tracers.
Thirty four cores have been chosen for mapping with same tracers on the basis of the brightness of the observed lines, especially $\rm N_2H^+$ line
to make follow-up mapping observations feasible, regardless of whether they are infall candidates or not. In this sense our sample is not 
biased in the selection of targets by the pre-existence of infall asymmetry in cores from the single pointing survey, but by the brightness of the sources.      

Therefore the overabundance of blue profiles in the  $\delta V$  distribution should be affected from our selection bias of the bright (in $\rm N_2H^+$) sources.
What is really implied in the  $\delta V$  distribution of Figure 1 would be that 
bright $\rm N_2H^+$ cores (i.e., cores with high column density) are likely to have inward gaseous motions.  
This result is further investigated in the following section. 

\subsection{Relation of the $\rm \delta V$ Distribution with Gas Column Density}
This section suggests that regions of higher column density have a better chance of having more negative $\rm \delta V$ or being in inward gaseous motion. 
We use the integrated intensity of $\rm N_2H^+$ 1-0 line as the column density tracer. The  $\rm N_2H^+$ 1-0 line is usually optically thin, and 
the ion molecule $\rm N_2H^+$ barely depletes in the dense core. Thus $\rm N_2H^+$ 1-0 is useful to trace the density distribution of the cores 
(e.g., Tafalla et al. 2004, 2006).

Figure 2 and 3 examine the dependence of $\rm \delta V$ on the integrated intensity of $\rm N_2H^+$ using single pointing data for CS 2-1, 
 3-2, HCN 1-0, and  $\rm N_2H^+$ from Lee et al. (1999, 2004) and Sohn et al. (2007).
The data were selected if the integrated intensity of $\rm N_2H^+$ is brighter than its 5 $\sigma$ uncertainty. This allows us to have 
47 sources for  $\rm \delta V_{CS 2-1}$, 44 sources for  $\rm \delta V_{CS3-2}$ and 37 sources for $\rm \delta V_{HCN 1-0}$. 
All the diagrams in Figure 2 and 3 show that the cores with stronger $\rm N_2H^+$ emission tend to have their profiles more blue-skewed.  

This tendency can be more thoroughly examined using mapping data in CS 2-1 in Figure 4.
A total number of the positions where both CS and  $\rm N_2H^+$ lines are available is 546, but in the figure 
we selected the data for 339 positions for which the integrated intensity of $\rm N_2H^+$ is brighter than its 5 $\sigma$ uncertainty.
We also display bigger red dots in Figure 4 which are the points of average values of $\rm \delta V$s and  $\rm N_2H^+$ intensities
for positions counted in order of increasing  $\rm N_2H^+$ intensity (i.e., from the left of the diagram) 
in 10 intervals so that in each interval the data for 34 positions are included and 
the rightmost bin contains last remainders (the data for 33 positions).
The error bar is 1 $\sigma$ uncertainty of the average. 
There is a clear tendency of $\rm \delta V$ to be larger in negative sense or of the infall asymmetry to be more significant
at the bright $\rm N_2H^+$ positions.

Figure 5 shows the distribution of $<\delta V>$ versus the peak integrated intensity of $\rm N_2H^+$ in the cores. 
Here $<\delta V_{CS}>$ is the average of $\delta V$ values of all positions for each core where the integrated intensity of 
$\rm N_2H^+$ is brighter than its 5$\sigma$ uncertainty. 
The difference between Figures 4 and 5 is that Figure 4 displays the distribution of $\delta V$ over all mapped positions while Figure 5
shows the distribution of average $\delta V$ over each core.
Figure 5 indicates that the infall asymmetry of CS line is more significant at a core with the brighter peak intensity of $\rm N_2H^+$. 

Figure 4 and 5 show that denser regions have bluer spectra. In addition, such dense regions have a larger proportion of blue spectra,
as shown in Figure 6.
The fractional blue excess  ($\rm E$) given in the ordinate of Figure 6 is defined as $\rm E = {(N_- - N_+)\over N_{tot}}$, 
where $\rm N_-$ is  the number of positions with
$\rm \delta V_{CS} \le -5~\sigma_{\delta V_{CS}}$, $\rm N_+$ is
the number of positions with $\rm \delta V_{CS} \ge 5~\sigma_{\delta V_{CS}}$, $\rm N_{tot}$ is the total number of positions given by 
$\rm  N_{tot} = N_- + N_+ +  N_0$, and $\rm N_0$ is the number of positions with 
$\rm  -5~\sigma_{\delta V_{CS}} < \delta V_{CS} <  5~\sigma_{\delta V_{CS}}$.  
All the numbers regarding the excess are estimated with the data points in the ten bins of the sample as 
in the case of Figure 4.
The figure again shows that the positions where  $\rm N_2H^+$ is brighter tend to have higher blue excess and 
thus to have more significant signature of inward motions. 
The  1 $\sigma$ uncertainty of the average in the $\rm N_2H^+$ intensity is typically less than 0.01 $\rm K~km~s^{-1}$ which is smaller than the size of 
the dots in the figure, except for that of the brightest data point which is about 0.07 $\rm K~km~s^{-1}$. 

The uncertainty of E is difficult to calculate from the noise of the spectra.
However, we may estimate its approximate value by assuming that the typical 1 $\sigma$ uncertainty in ($\rm N_- - N_+$)  is unity, 
corresponding approximately to one incorrect assignment in the sample of 
$\rm N_{tot}=34$ in each bin.  This implies by standard propagation of errors, uncertainty of E of $\rm {1 \over N_{tot}} \approx 0.03$.
This suggests that the uncertainty of E is fairly small like the case of the uncertainty of the $\rm N_2H^+$ intensity.
Thus the apparent trend in Figure 6 is likely to be significant against uncertainties of the $\rm N_2H^+$ intensity and of E.

\subsection{Variation of  $\rm \delta V_{CS}$ with the Distance from the Peak Gas Column Density}
This section examines how the asymmetric pattern of spectral lines changes with the distance from the density peak of a core. 
    
We discuss this with Figure 7 where $\rm \delta V_{CS}$ of each spectrum is plotted against the distance of its position  
from the peak intensity position of $\rm N_2H^+$.
The distances to the cores from us were adopted from Table 2 of Lee \& Myers (1999).
Figure 7(a) displays $\rm \delta V_{CS}$ distribution of the spectra of all sources where the integrated intensity of $\rm N_2H^+$ is 
brighter than its 5 $\sigma$ uncertainty, indicating that $\rm \delta V_{CS}$ tends to be more negative at positions closer to the position of peak intensity.
In the figure there is a break for this tendency at the radius of  $\sim$ 0.1 pc, where the local mean value of $\rm \delta V_{CS}$ is close to zero. 
Over 0.1 pc the $\rm \delta V_{CS}$ becomes again negative, although it is less negative than at inner 
($<\sim$0.07 pc) positions, implying that blue profiles are also prevalent even at large radii.
The break around  $\sim$ 0.1 pc is mainly due to the contribution by the data from three cores  L183, L429-1, and L1495A-N 
where both blue and red asymmetric CS profiles are observed in a significant number.

Figure 7(b) is the same diagram as Figure 7(a), but without having these cores, demonstrating that the mean value of $\rm \delta V_{CS}$ 
close to zero in Figure 7(a) is due to the distribution of blue and red profiles in a comparable number in three cores.  
L1512 is also another core showing such distribution of spectra. But it has a relatively small number (9) of data positions and 
its most positions in the diagram are located within the radius of 0.06 pc from the  $\rm N_2H^+$ peak position.
Thus L1512 is not responsible for the break at  $\sim$ 0.1 pc.    

Figure 8 is another display of the $\rm \delta V_{CS}$ distribution with the distance from the position of the peak $\rm N_2H^+$ intensity
using a tool of the fractional blue excess like Figure 6.
The figure shows that the fractional blue excess is more pronounced, implying overabundance of blue profiles, in inner region  of the core (within 
a radius of about 0.07 pc from the column density peak of the core) than in its outer region. 
The  1 $\sigma$ uncertainty of the average in the radial distance from the position of the peak $\rm N_2H^+$ intensity is typically less than 
0.003 pc which is smaller than the size of
the dots in the figure, except for that of the most distant point which is 0.017 pc.
Because the uncertainty of E is also expected to be fairly small as discussed in the last section,
this trend seen in Figure 8 is expected to be significant.

As the case of Figure 7, Figure 8(a) is a diagram for the spectra of all sources, showing that blue excess is close to zero at $\sim$ 0.1 pc and 
become significant even at the radii larger than 0.1 pc.
Figure 8(b) is the same diagram as Figure 8(a), but without above three cores, confirming that the break of the blue excess at $\sim$ 0.1 pc is due to these cores. 

The foregoing figures show that most cores in our sample have spectral asymmetry maps indicating contracting motions.
However it is also clear that a minority of the cores cannot be described as primarily contracting.
Therefore we classify cores into four groups according to their apparent dynamical status. 
We then examine how $\rm \delta V_{CS}$ varies with the distance from the core map peak within each group.
 
The classification of the cores was made by using two parameters given by LMT01. These are the fractional blue excess E and the P-value of a
student t-test for the  $\rm \delta V_{CS}$ distribution for each core. The P-value is the probability of drawing our  $\rm \delta V_{CS}$ 
distribution from a zero mean t-distribution. 
The E parameter determines how many blue or red profiles exist in a core while the P parameter informs how the dominance of the blue or red profiles is 
significant in the core. These parameters are listed in Table 3 of LMT01.  
Here the classification of the cores is given for 33 cores with $\rm \delta V_{CS}$ measurement for more than three positions.

The first group of the sources are the infall candidates suggested by LMT01 which show significant overabundance of blue profiles in a core
($\rm E \ga 0.10$ and $\rm P \la  0.1$). This corresponds to the groups 1 and 2 in Figure 6 of LMT01; 
L1355, L1498, L1495A-S, TMC2, L1544, TMC1, L1552, L1622A-2, L158, L183,
L1689B, L234E-C, L234E-S, L492, L694-2, L1155C-2, L1155C-1, L981-1, and L1197. Here we name these as candidates of ``contracting core''. 

The second group of the sources is the cores where blue and red profiles are observed in a comparable number so that there is no significant 
blue or red excess ($E\approx 0$) with a large spread in the $\delta V_{CS}$ distribution (one standard deviation $\ga 0.4$).
This corresponds to group 5 in Figure 6 of LMT01; L1495A-N, L1507A, and L1512. We refer this group to candidates of ``oscillating core'' 
because the mixture of blue and red profiles may be caused by oscillation motions of gas as suggested  by  Lada et al. (2003), 
Redman et al. (2006), and Aguti et al. (2007).

The third group of the sources is the cores dominated by red asymmetric profiles, showing significant overabundance of red profiles in a core
($\rm E \la -0.15$ and $\rm P \approx 0.0$). This corresponds to the group 3 in Figure 6 of LMT01; L134A,  L429-1, and CB246. 
L134A has been dropped from this group in LMT01 by a mistake 
because it was erroneously treated as the sources with a small number ($< 7$) of $\rm \delta V_{CS}$ measurements although it had 9 positions 
where $\rm \delta V_{CS}$ was obtained. Now we include L134A in this group. 
On the other hand L1521F which had been classified to this group in LMT01 is dropped from this group  because it is now known 
to have an embedded source (Bourke et al. 2006).
We refer this group to the candidates of ``expanding core''.

The fourth group of the sources are the cores with little blue or red excess and CS profile shapes similar to a single Gaussian form
so that there is no significant blue or red excess ($E\approx 0$) with a small spread in $\delta V_{CS}$ distribution (one standard deviation $\la 0.3$). 
This corresponds to the group 4 in Figure 6 of LMT01; L1333, L1495B, L1400A, CB23, L1517B, L1622A-1, L1696A, and L234E-N.
Here we included to this group CB23 which had been dropped 
from any classification of the cores in LMT01 because of a small number (four) of measurements of $\rm \delta V_{CS}$.
We refer this group to the candidates of ``static core''. 

Figure 9(a) displays the $\rm \delta V_{CS}$ distribution versus the distance from the position of $\rm N_2H^+$ peak intensity 
for 19 contracting cores in group 1. In this group, there are more positions with negative  $\rm \delta V_{CS}$ than those with positive  $\rm \delta V_{CS}$
at every radius. Within $\sim$0.07 pc this trend is much more significant: nearly all of the positions in contracting cores 
have negative $\rm \delta V_{CS}$. 

On the other hand, the trend is not seen in other groups of the cores. 
Figure 9(b) shows that the oscillating cores in group 2 have overall comparable positions at both positive 
and negative  $\rm \delta V_{CS}$ along the distance, 
although there are a few more positions with negative $\rm \delta V_{CS}$ at inner region $< 0.04$ pc and more
positions with positive $\rm \delta V_{CS}$  at outer region  $> 0.1$ pc.
The expanding cores in group 3 show opposite distribution to that of contracting cores, i.e., much more positions with positive $\rm \delta V_{CS}$ 
than those with negative  $\rm \delta V_{CS}$ [Figure 9(c)].
The static cores in group 4 show no significant variation in $\rm \delta V_{CS}$ value along the distance from the peak $\rm N_2H^+$ intensity, but with one exceptional 
position with the highest positive $\rm \delta V_{CS}$ at the longest distance. 

Note that the contracting cores are the majority of the data and their group shows the similar $\rm \delta V_{CS}$ distribution to what is found in Figure 7 and 8
for whole sample of cores while other groups of the cores do not show it. This suggests that 
the characteristic property of the predominant $\rm \delta V_{CS}$ distribution 
of negative value along the radius of the cores comes from  $\rm \delta V_{CS}$ distribution in the contracting cores.

\section{Discussion }

\subsection{Relation of the $\rm \delta V$ Distribution with Gas Column Density and its Implication}
In Section 3.2 we have shown that the bright  $\rm N_2H^+$ positions tend to have more negative  $\rm \delta V_{CS}$,
a significant signature of inward motions. This section discusses this result in terms of individual cores and their dynamical status, and
interprets the discussion in terms of evolution of the dense core.

Figure 10 indicates the distribution of $\rm \delta V_{CS}$ against the integrated intensity of $\rm N_2H^+$
for all available positions in the starless cores in four panels. 
While Figure 4 presents these data for all the cores in the plot, Figure 10 shows the same data in a separate plot for each core. 
Panel (a) of Figure 10 displays the $\rm \delta V_{CS}$ distribution against the integrated intensity of $\rm N_2H^+$ for the contracting cores in 
decreasing order of $\rm -<\delta V_{CS}>$, showing that $\rm \delta V_{CS}$ clearly tends to be more negative at the 
brighter positions of  $\rm N_2H^+$.
On the other hand, the cores in the other groups do not show any trend with integrated intensity of $\rm N_2H^+$ 1-0.
The oscillating cores in panel (b) show that the distribution of  $\rm \delta V_{CS}$ is polarized between positive and negative values of  
$\rm \delta V_{CS}$,  independent of the integrated intensity of $\rm N_2H^+$. 
The expanding cores in panel (c) are also hard to see any dependency of the distribution of  $\rm \delta V_{CS}$ 
on the integrated intensity of $\rm N_2H^+$. Instead they have mostly positive $\rm \delta V_{CS}$. 
The static cores in panel (d) tend to be less bright in  $\rm N_2H^+$ emission than other types of the cores and show a uniform small spread of  $\rm \delta V_{CS}$
compared with that of oscillating cores.  

Figure 11 is another presentation of the distribution of $<\delta V_{CS}>$ 
as a function of the peak integrated intensity of $\rm N_2H^+$ when the cores are combined into their four dynamical status groups.

Figure 11(a) displays that contracting cores have the distribution of negative $<\delta V_{CS}>$ over wide range of the peak 
integrated intensity of $\rm N_2H^+$.
On the other hand, the cores in other groups have the $<\delta V_{CS}>$ distribution in smaller range of the intensity of $\rm N_2H^+$.
The oscillating cores and the static cores are similarly distributed near to zero $<\delta V_{CS}>$ value, but 
the oscillating cores have larger scatter (s.e.m. of 0.13 - 0.18) in $<\delta V_{CS}>$ value of each core because of the mixture of blue and red profiles in the core 
[Figure 11(b)], compared with that (s.e.m. $< 0.1$ mostly) in  $<\delta V_{CS}>$ value of static cores [Figure 11(d)].
All of the expanding cores have positive $<\delta V_{CS}>$ contrary to the $<\delta V_{CS}>$ distribution of contracting cores.
  
Figure 11 shows that brighter cores tend to be dominated by contracting cores as was also shown in Figure 6. In addition, 
Figure 11 shows that non-contracting cores tend to be similarly faint in each group, with  $\rm \int  T^*_A(N_2H^+) dv \la 1~K~km~s^{-1}$.
Above a certain value ($\rm \sim 1.6~K~km~s^{-1}$) of the integrated intensity of $\rm N_2H^+$ 
there are contracting cores only (with negative $<\delta V_{CS}>$). This suggests that as starless cores become denser and 
approach some specific column density, they tend to have more chance to contract. Above this critical value of column density all starless cores 
are always to be in inward motions. 
We derived such a critical column density by using Equation (A4) of Caselli et al. (2002) and 
converted this to the corresponding $\rm H_2$ column density of $\rm 6\times 10^{21}~cm^{-2}$  
by using an average abundance of  $\rm N_2H^+$ of $\sim 6.8(\pm 4.8)\times 10^{-10}$ of starless cores
which is obtained using Table 3 in Johnstone et al. (2010). 
Our diagram may indicate that there is a threshold value of column density of  $\sim \rm 6\times 10^{21}~cm^{-2}$ over which
most of starless cores are contracting. 
Note that this is very similar to the value  of column density of $\rm 8.0\times 10^{21}~cm^{-2}$  
for a star formation in dense cores suggested by Onishi et al. (1998)

Crapsi et al (2005) have shown that $\rm N_2H^+$ column density [$\rm N(N_2H^+)$] can be an indicator of 
the evolutionary status of starless cores since $\rm N(N_2H^+)$ increases with other evolutionary indicators including 
$\rm N_2D^+$ column density [$\rm N(N_2D^+)$], $\rm N(N_2H^+)$/ $\rm N(N_2D^+)$, $\rm H_2$ column density, and CO depletion factor.
According to their investigation, it is likely that starless cores with higher $\rm N_2H^+$ column density are more evolved 
than the cores with lower $\rm N_2H^+$ column density.
If this is true,  Figure 11 may imply that, in terms of increasing column density, starless cores in four groups are 
in a sequence of core evolution. In this picture, the static cores with the lowest column density may be in the earliest stage,
the expanding cores and/or the oscillating cores in the next, and the contracting cores (with the highest column density) in the latest stage.
We note that oscillating and expanding cores are nearly indistinguishable because of a small number of samples.
It may be possible that some expanding cores are at the stage of outward motion in oscillatory mode.   
More sample of the cores in each stage is needed to constrain the evolutionary sequence of the cores with better statistical significance.

The discussion for this evolutionary sequence is qualitatively similar to the picture discussed by  Stahler \& Yen (2010).
Using perturbation theory, they showed that the starless core would undergo contraction if they are
initially compressed or inflated by the oscillation. Our diagram of observing data seems to indicate a similar result:
Once the static cores are perturbed, this may result in imbalance between self gravity and the outward pressure gradient, and thus
expanding or oscillating motions in the core.
Once a core becomes denser over the critical column density like $\rm 6\times 10^{21}~cm^{-2}$, it may be in ``all'' contracting motions. 
We note that there are also cores with low column density, but in inward motions, indicating that some cores may spend very short time in 
oscillation or expansion. 

The discussion for the time scale in each stage is also interesting.
Lee \& Myers (1999) have suggested that starless cores of a mean density of $\rm 6-8 \times 10^3~cm^3$ last for about
$\rm 0.3 - 1.6 \times 10^6$ years using the statistical number ratio ($\sim 3.26$ for the number ratio 
of starless cores and cores with embedded protostars) and the life time of protostar as $\rm 1-5 \times 10^6$ years. 
Ward-Thompson et al. (2007) re-estimated this life time of starless cores as $\sim 6 \times 10^5$ years using better estimate of Class I life time
($\sim 2 \times 10^5$).
This estimation also needs to be revised by considering that 
about 20\% of the starless cores may not be ``starless'' because they contain an embedded very faint source, i.e., the VeLLO (Dunham et al. 2008) 
and the reference life time of protostars (Class 0 and I) is recently updated as $\rm 0.54 \times 10^6$ years with much better statistics 
by Evans et al (2009). By using these new information, our revised value of the life time of the detectable starless cores denser 
than $\rm \sim 10^4~cm^{-3}$ is now estimated to be $\rm \sim 1.4 \times 10^6$ years. 

If the starless cores share this duration for each evolutionary stage proportional to the statistics of the numbers in each class, 
the cores may spend the force balanced period for  $\sim 3 \times 10^5$ ($\approx{8 \over33} \times 1.4 \times 10^6$) years,  experience 
either expanding or oscillatory motions in the surface of the cores
for $\sim 3 \times 10^5$ ($\approx {6 \over33} \times 1.4 \times 10^6$) years and then contract 
for  $\sim 8 \times 10^5$ ($\approx {19 \over33} \times 1.4 \times 10^6$) years.

Aguti et al. (2007) have calculated the modes of pulsation for an isothermal globule of gas in spherical shape for the core 
FeSt 1-457 and found its required mode as the least-damped mode of $l=2$ with an oscillation period of $\sim 3\times 10^5$ years.
This oscillation period is consistent with our determination of the statistical period of starless cores in oscillation mode,
even with small sample statistics.  

\subsection{Environmental Effect on Internal Motions in Starless Cores}
Are internal motions in starless cores affected by the core environment ?
We examined the association of  35  $\rm N_2H^+$ cores (including B68 and Fest1-457 as well as our sample) to any surrounding structures of clouds in the $\rm A_v$ maps 
of the Digitized Sky Survey obtained using a star-count method by Dobashi et al. (2005), by checking how the core is enclosed 
with the lowest contour ($\rm A_v$=0.5) in the $\rm A_v$ maps.
 
From this eye inspection we found that among 35  $\rm N_2H^+$ cores L694-2 is a core isolated entirely from nearby large cloud complex.
There is also one small cloud where 2 of the cores, L134A and L183, are enclosed within the lowest contour ($\rm A_v$=0.5). 
This cloud is isolated from other nearby clouds.
B68 is not identified in the $\rm A_v$ map by Dobashi et al. (2005),  probably
because there are not enough background stars available toward B68 for enabling $\rm A_v$ around the core to be estimated.
From its location in the $\rm A_v$ map, B68 seems located in a very small $\rm A_v$ region within a surrounding cloud.
Thus we believe 4 sources (L694-2, L134A, L183, and B68) may be well isolated from large clouds.

On the other hand the majority of the cores (17 of 33 cores) are found to exist as a part of long filament clouds. 
They are L1333, L1498,  L1495B,  L1495A-N,  L1495A-S,  L1400A, TMC2, TMC1, L1507A, CB23, 
L1544, L1696A, L1689B, L158, L234E-N, L234E-C, and L234E-S.
The rest of our sample (14 of 35 cores) are more likely surrounded in a large cloud or located in the edge of the cloud 
(L1355, L1517B,  L1512, L1552,  L1622A-2, L1622A-1, L492, L429-1, L1155C-2,  L1155C-1 L981-1, L1063, L1197, CB246, and Fest1-457).  

Can isolation of the cores from a large complex cloud affect its dynamical status ? 
We may find some clue for this question if the isolated cores have their characteristic features of the spectral asymmetry.
According to criteria given in Section 3.3, L694-2 and L183 are classified as contracting cores and L134A  as expanding core.
B68 is known to have a mixture of blue and red profiles, interpreted as 
an indication of oscillation motions of gaseous material in the surface of a core (Redman et al. 2006, Lada et al. 2003).  
It is interesting to note that no core which is in isolation from other clouds has spectra of a  Gaussian shape indicative of
dynamically static status.  
However, we are cautious to draw any conclusive remarks on the characteristic properties 
of the dynamical status of the isolated cores due to the poor statistics by our small sample and more data are needed.  

Now we examine difference in dynamical status of starless cores in two other environments, the long filament clouds and the large complex clouds.
According to previous classification criteria of the cores and examination of core associations with large clouds, we find that 
nine contracting, two oscillating, and six static cores are located in the  long filamentary clouds 
while eight contracting, one oscillating, two expanding cores, and three static cores are in large complex clouds.
Although there is a slight difference in the number of cores in two types of clouds, 
it seems hard to conclude whether the difference is significant or not, due to its limited statistics.  


\section{Summary}
This paper aims for discussing how it is likely for a starless core to be in contracting, expanding, 
oscillating or static motions 
and which conditions are related to such dynamical status of the core, by analyzing asymmetric pattern in molecular lines.
For this purpose we collected either single-pointing or mapped spectroscopic line data for several tens of starless cores 
and re-analyzed those especially by using the normalized velocity difference  $\rm \delta V_{CS}$ between optically thick and thin lines.

The main conclusions we found from this analysis are as follows:

1. Blue profiles are dominant in starless cores, implying that inward motions are prevalent in starless cores.
These profiles are found to be more abundant and their asymmetry is bluer 
at the positions of the core with stronger $\rm N_2H^+$ or higher column density.
This indicates that positions with high column density  in the core are more likely contracting.

2. Relying on the distribution of their spectral asymmetric features, 
starless cores were classified to have four types of motions,
contracting, oscillating, expanding, and static motions.
More than half (19) of the 33 cores studied have spectral line maps dominated by evidence of contracting motions, 
about a quarter (8) of the cores show no significant evidence of either contraction or expansion,  
about 10\% (3) of the cores are dominated by expanding motions, and about  10\% (3) show a significant mixture of contracting 
and expanding motions, which may be interpreted as oscillatory motions. 

3. Most of the contracting cores tend to have more positions with negative  $\rm \delta V_{CS}$ (blue asymmetry) than positive  $\rm \delta V_{CS}$ 
at every radius, with more concentration of negative $\rm \delta V_{CS}$ within $\sim$0.07 pc from peak  $\rm N_2H^+$ intensity region.
This implies that inward motions in contracting cores are occurring along most lines of sight, with predominance of those motions 
near the column density peak region of the cores.

4. Contracting cores are far more numerous for the bright cores. Above a certain peak column density ($\sim \rm 6\times 10^{21}~cm^{-2}$) 
only contracting cores are seen and no oscillating, expanding, or static cores are seen. 
This implies that as starless cores become denser, 
they are more likely to contract. Above some critical value of column density all starless cores
are likely to be contracting.
 
5. In terms of increasing column density, starless cores in different internal motions may reflect their different status of evolution: 
static cores in the earliest stage, expanding and/or oscillating cores in the next, and the contracting cores in the latest stage.
This is consistent with a theoretical picture by Stahler \& Yen (2010) where once a static cores is perturbed, the core would show 
expanding or oscillating motions, and would then begin prolonged contraction motions.
On the other hand, we note that some starless cores have both low column density and inward motions.
This may mean that these cores may not spend much time in oscillation or expansion.


\acknowledgments 
This research was supported by Basic Science Research Program though the National Research Foundation of Korea (NRF) funded by the Ministry of 
Education, Science and Technology (2010-0011605). 

%
%
\clearpage






\clearpage
\begin{figure}
\centering
\includegraphics[height=7.5in,angle=0]{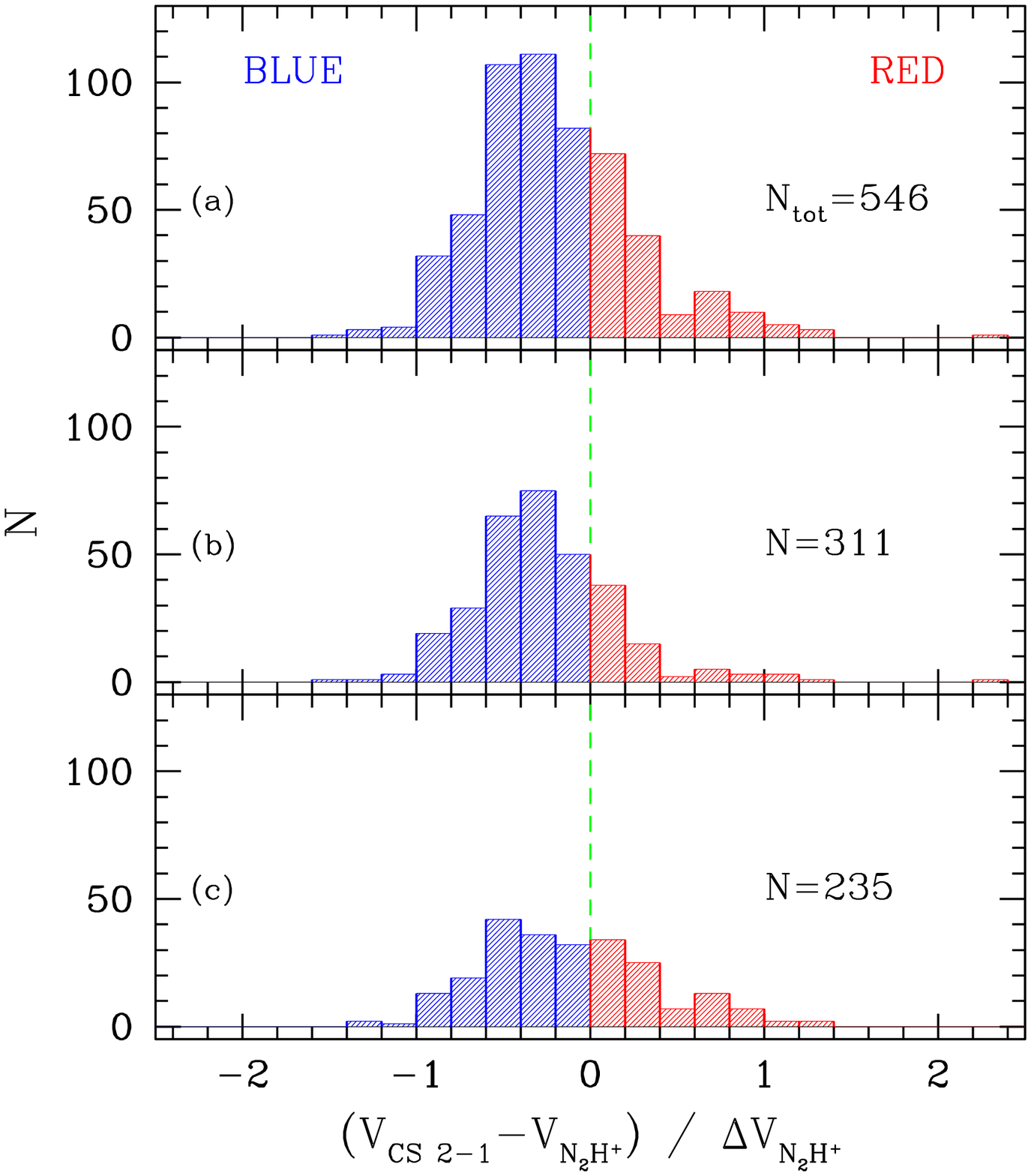}
\caption{Number distribution of $\rm \delta V_{CS}$ for 34 starless $\rm N_2H^+$ cores.
Upper panel is a histogram of the number distribution of  $\delta V$ for all 546 positions of 34 starless  
$\rm N_2H^+$ cores, middle panel is that for the cores selected as the infall candidates 
from a single pointing survey, and lower panel is that of the cores selected based on strong detectability with high density tracers
or optical opaqueness.
}
\end{figure}

\clearpage
\begin{figure}
\centering
\includegraphics[height=7.5in,angle=270]{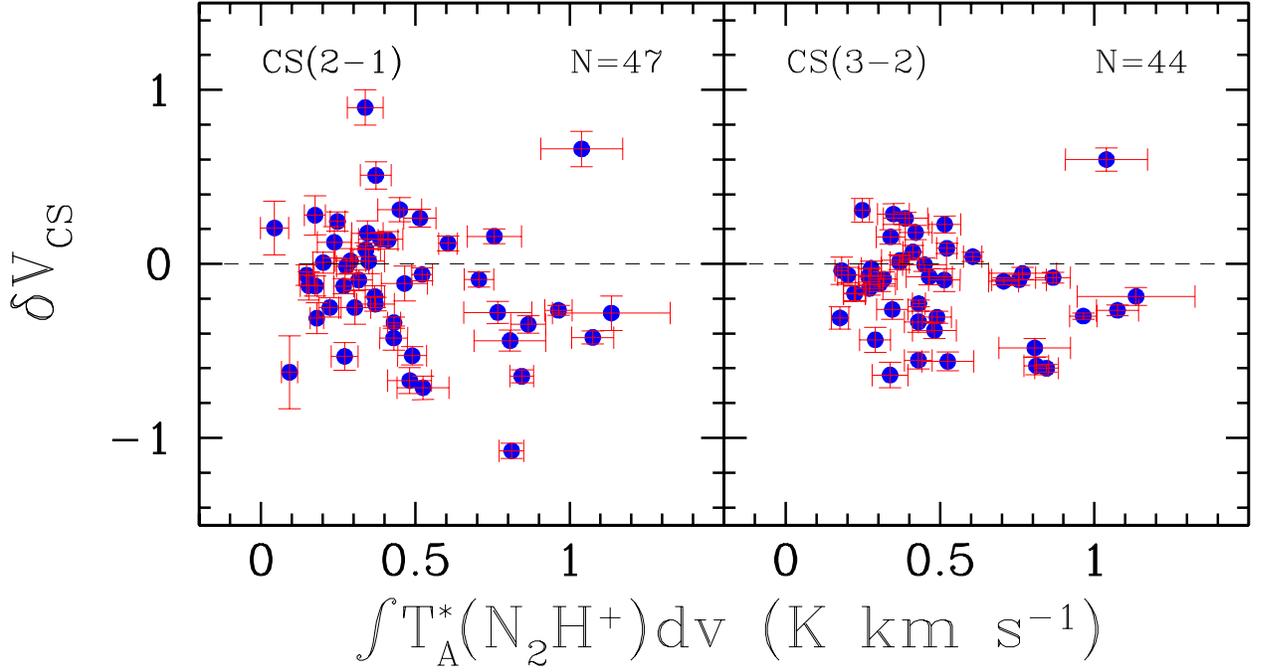}
\caption{
$\rm \delta V_{CS}$ distribution for CS 2-1 and CS 3-2 vs. the integrated intensity of $\rm N_2H^+$
for cores from the single pointing survey of LMT99 and LMP04.
The data for which the integrated intensity of $\rm N_2H^+$ is brighter than its 5 $\sigma$ are used in the diagram.
Note that all sources for which the integrated intensity of $\rm N_2H^+$ is stronger than $\sim 0.7$ $\rm K~km~s^{-1}$
have negative $\rm \delta V_{CS}$ except for L63.
}
\end{figure}

\clearpage
\begin{figure}
\centering
\includegraphics[height=7.5in,angle=270]{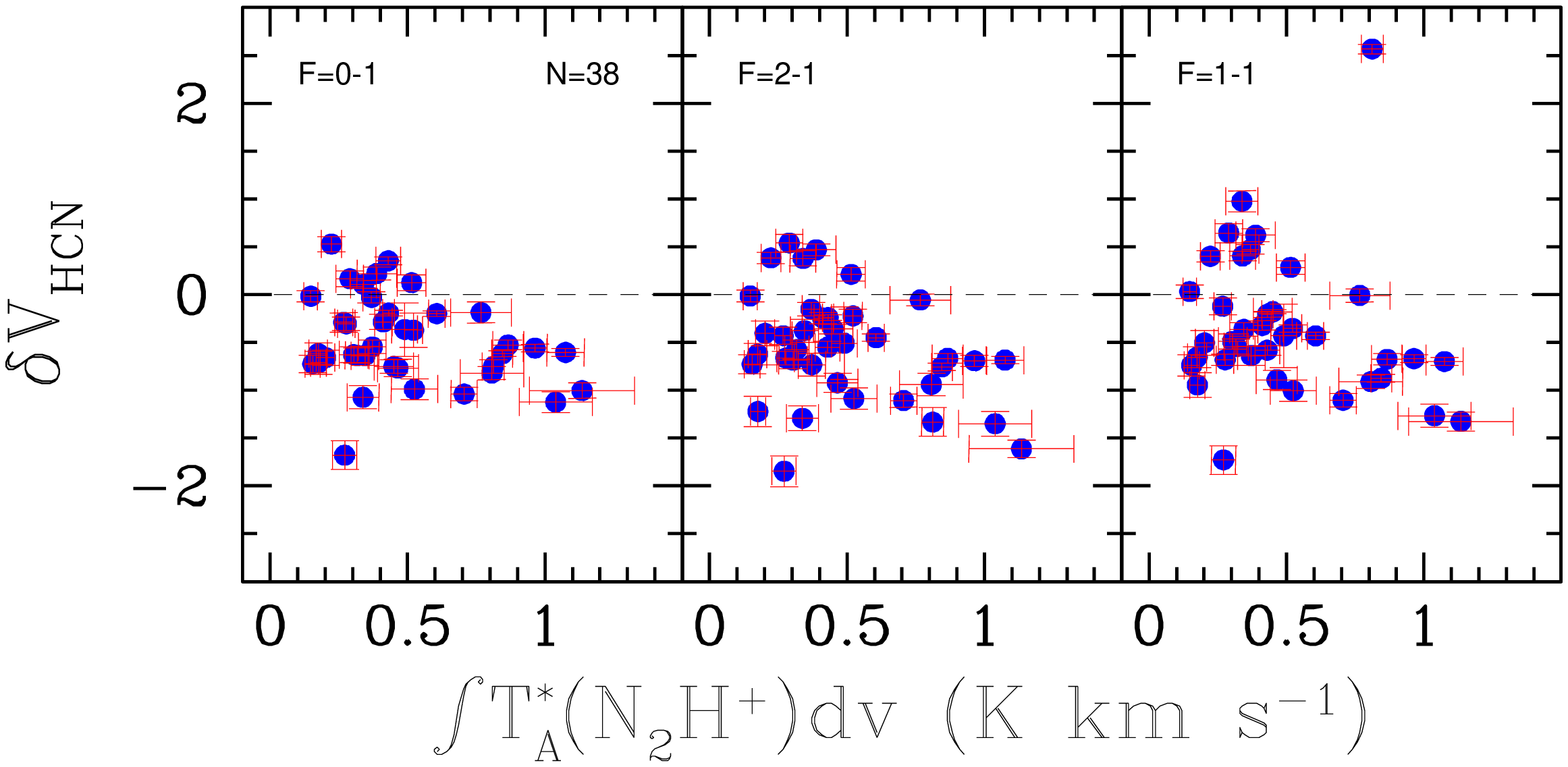}
\caption{
$\rm \delta V_{HCN}$ distribution vs. the integrated intensity of $\rm N_2H^+$ for 39 cores from the single-pointing surveys of Sohn et al. (2007).
The data for which the integrated intensity of $\rm N_2H^+$ is brighter than its 5 $\sigma$ are used in the diagram.
Note that all sources for which the integrated intensity of $\rm N_2H^+$ is stronger than $\rm \sim 0.6~K~km~s^{-1}$
have negative $\rm \delta V_{CS}$ except for L183 in HCN 1-0 F=1-1.  
}
\end{figure}

\clearpage
\begin{figure}
\centering
\includegraphics[width=7.5in]{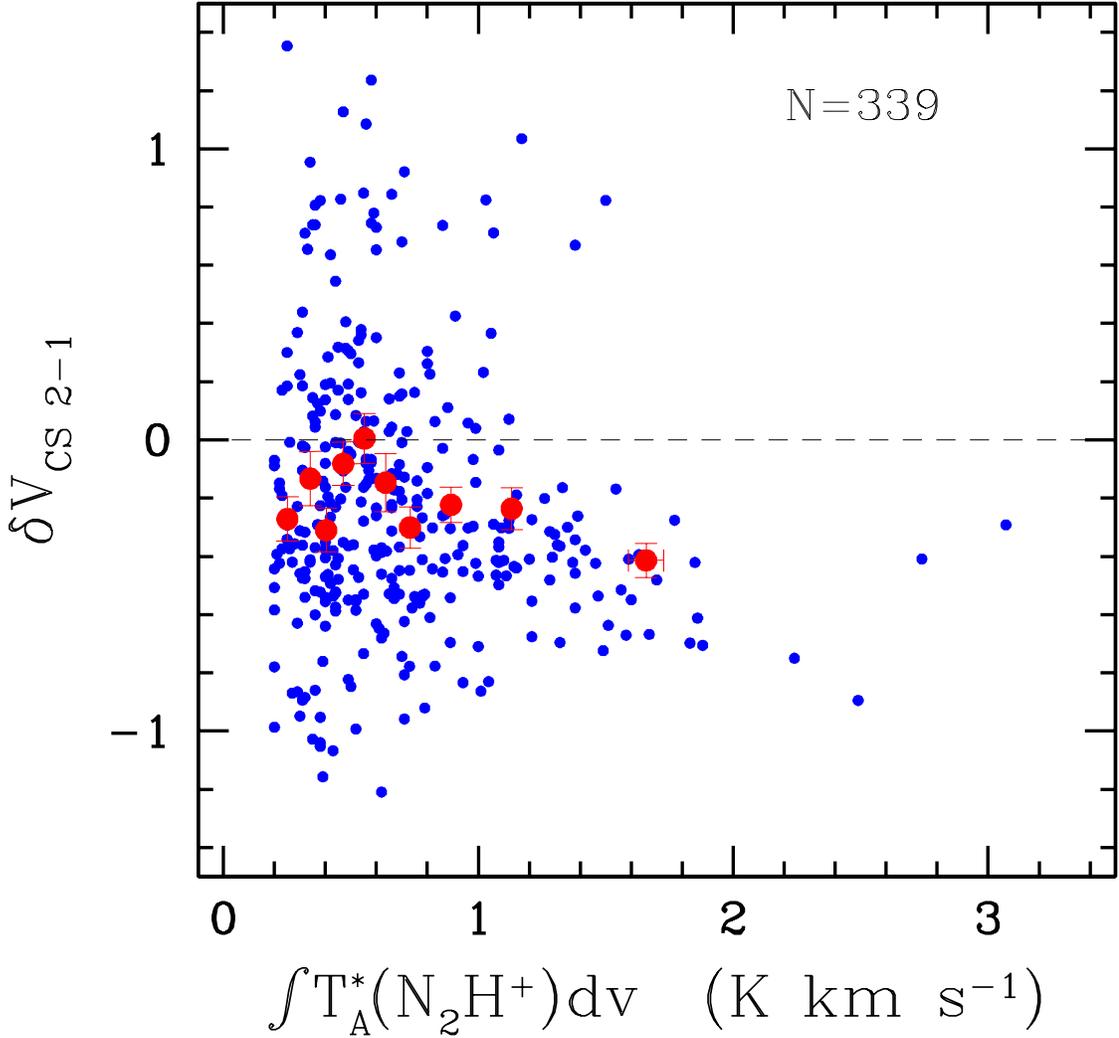}
\caption{
$\rm \delta V_{CS 2-1}$ distribution vs. the integrated intensity of $\rm N_2H^+$ for 34 cores from the mapping studies of LMT01.
The data for which the integrated intensity of $\rm N_2H^+$ is brighter than 5 $\sigma$ are used in the diagram.
Every bigger red dots are the points of average values of $\rm \delta V$s and  $\rm N_2H^+$ intensities
for positions counted in order of increasing $\rm N_2H^+$ intensity in 10 intervals.
An error bar is 1 $\sigma$ of the average.
Note that in the figure we have one position which has a  $\rm \delta V_{CS2-1}$ of 2.247, out of the Y-axis range of the figure.
}
\end{figure}

\clearpage
\begin{figure}
\centering
\includegraphics[width=7.5in]{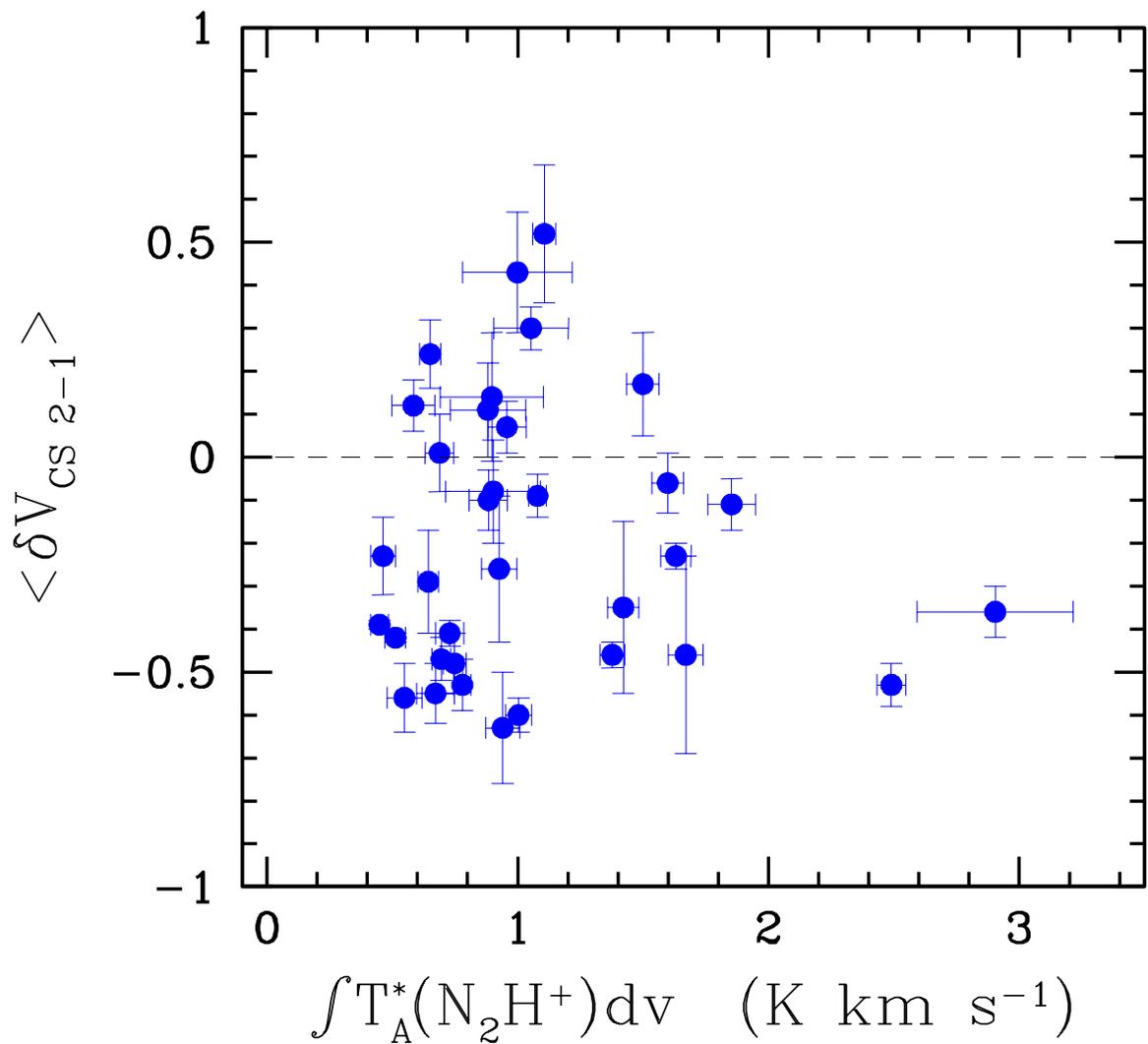}
\caption{
$\delta V$ distribution vs. the peak integrated intensity of $\rm N_2H^+$ within each starless core.
$\rm <\delta V_{CS}>$ is the average of $\delta V$s for all positions of each source where integrated intensity of
$\rm N_2H^+$ is brighter than its 5$\sigma$.
}
\end{figure}

\clearpage
\begin{figure}
\centering
\includegraphics[width=7.5in]{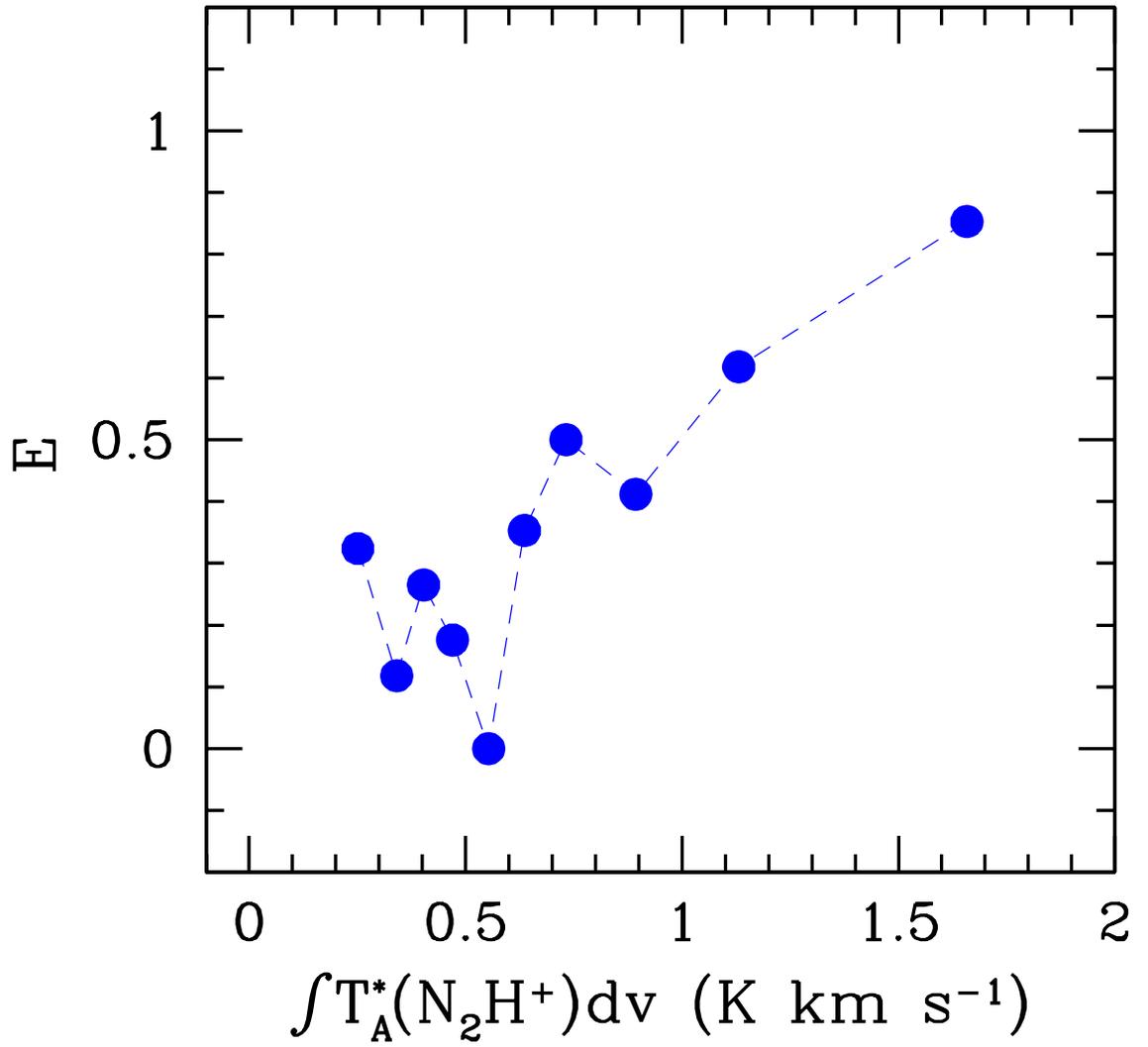}
\caption{
Fractional blue excess against the integrated intensity of $\rm N_2H^+$.
The excesses were estimated with the data points in the 10 bins of the sample as
the case of Figure 4. 
}
\end{figure}

\clearpage
\begin{figure}
\centering
\includegraphics[width=5.5in,angle=270]{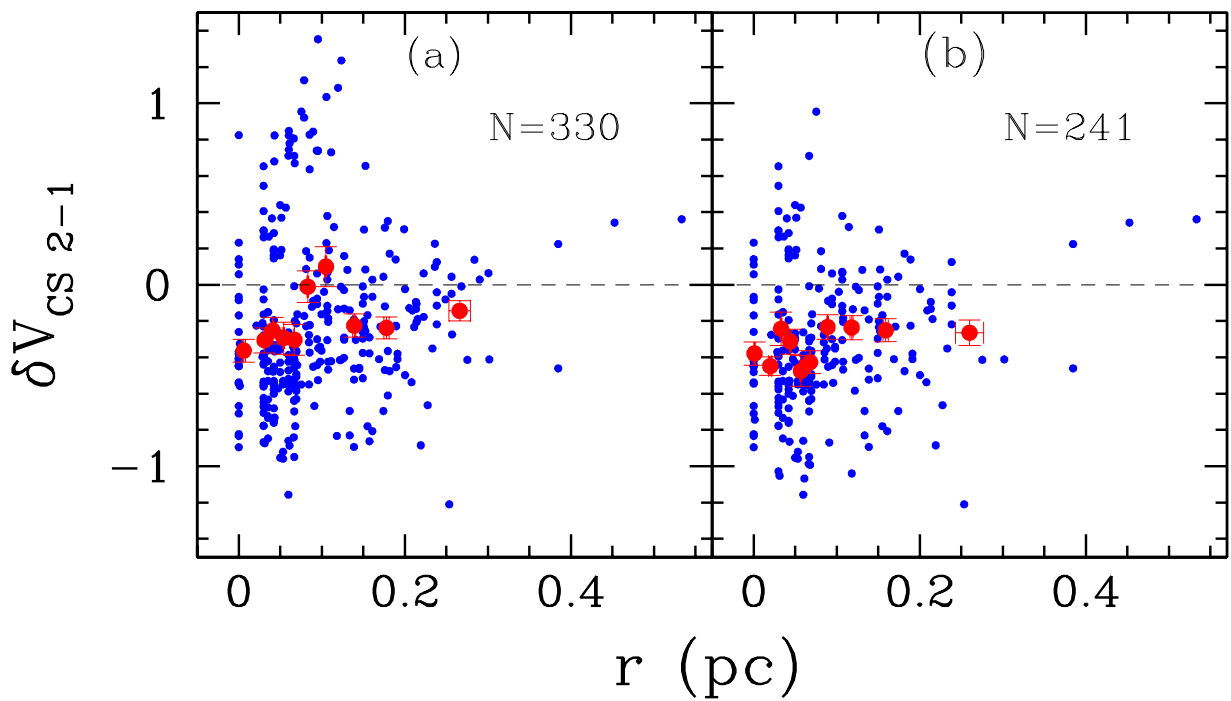}
\caption{
$\rm \delta V_{CS}$ distribution against the distance from the peak intensity of $\rm N_2H^+$ of the cores.
Panel (a) is the $\rm \delta V_{CS}$ distribution of the spectra of all sources where the integrated intensity of $\rm N_2H^+$ is
brighter than its 5 $\sigma$ uncertainty 
while panel (b) is the same diagram as the panel (a), but without three cores, L183, L429-1, and L1495A-N 
where both blue and red asymmetric CS profiles are seen in a comparable number.
}
\end{figure}

\clearpage
\begin{figure}
\centering
\includegraphics[width=5.5in,angle=270]{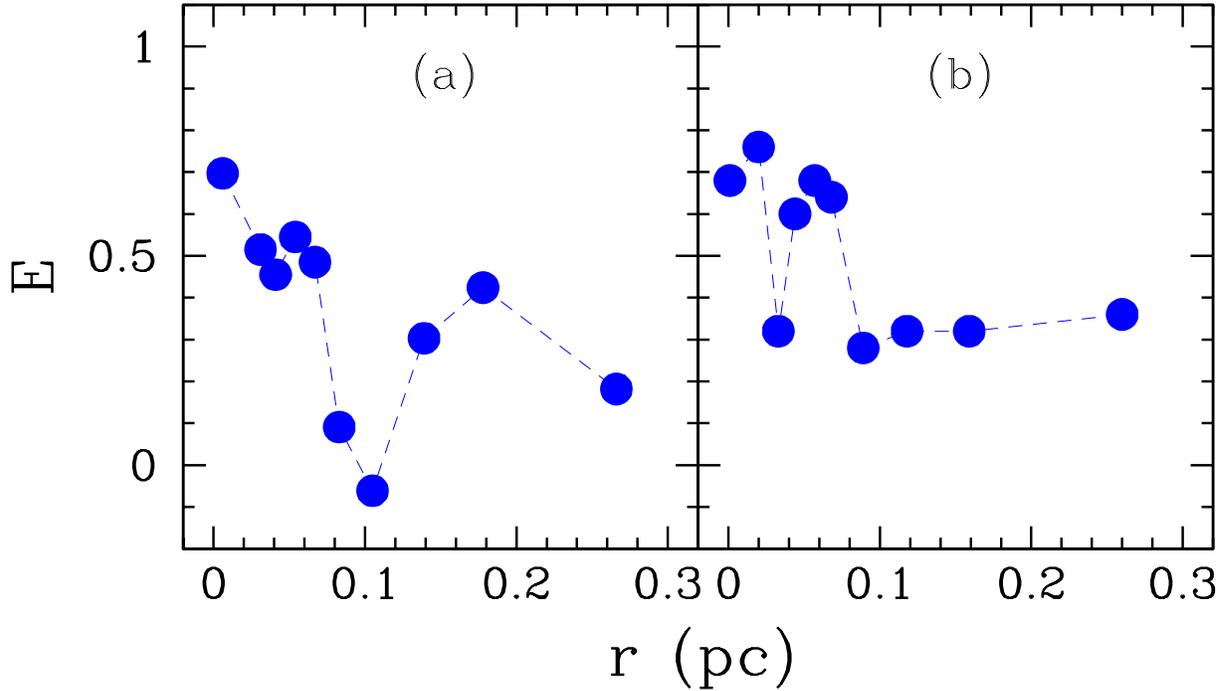}
\caption{
Fractional blue excess against the distance from the peak intensity of $\rm N_2H^+$ of the cores.
Like Figure 7, panel (a) is a diagram for the spectra of all sources while panel (b) is the same as panel (a), but without three cores
L183, L429-1, and L1495A-N where  both blue and red asymmetric CS profiles are seen in a comparable number.
Both diagrams indicate that the blue excess is the highest near the peak intensity of  $\rm N_2H^+$ and
significant  even at the radii larger than 0.1 pc.
}
\end{figure}

\clearpage
\begin{figure}
\centering
\includegraphics[height=7.5in,angle=0]{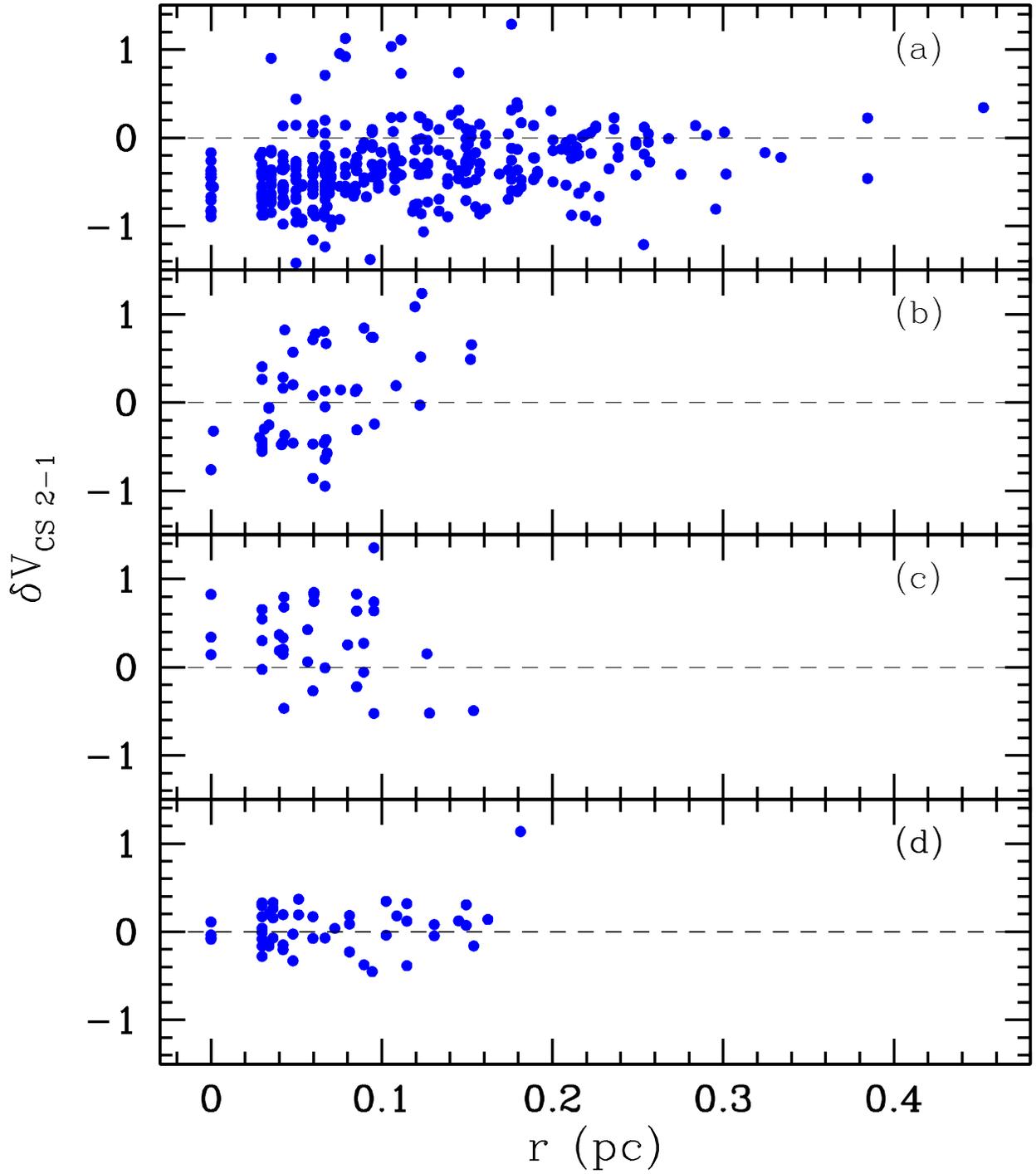}
\caption{
$\rm \delta V_{CS}$ distribution against the distance from a position of the peak $\rm N_2H^+$ intensity for 
cores in (a) contracting, (b) oscillating, (c) expanding, and (d) static motions.
}
\end{figure}

\clearpage
\begin{figure}
\centering
\includegraphics[width=7.5in]{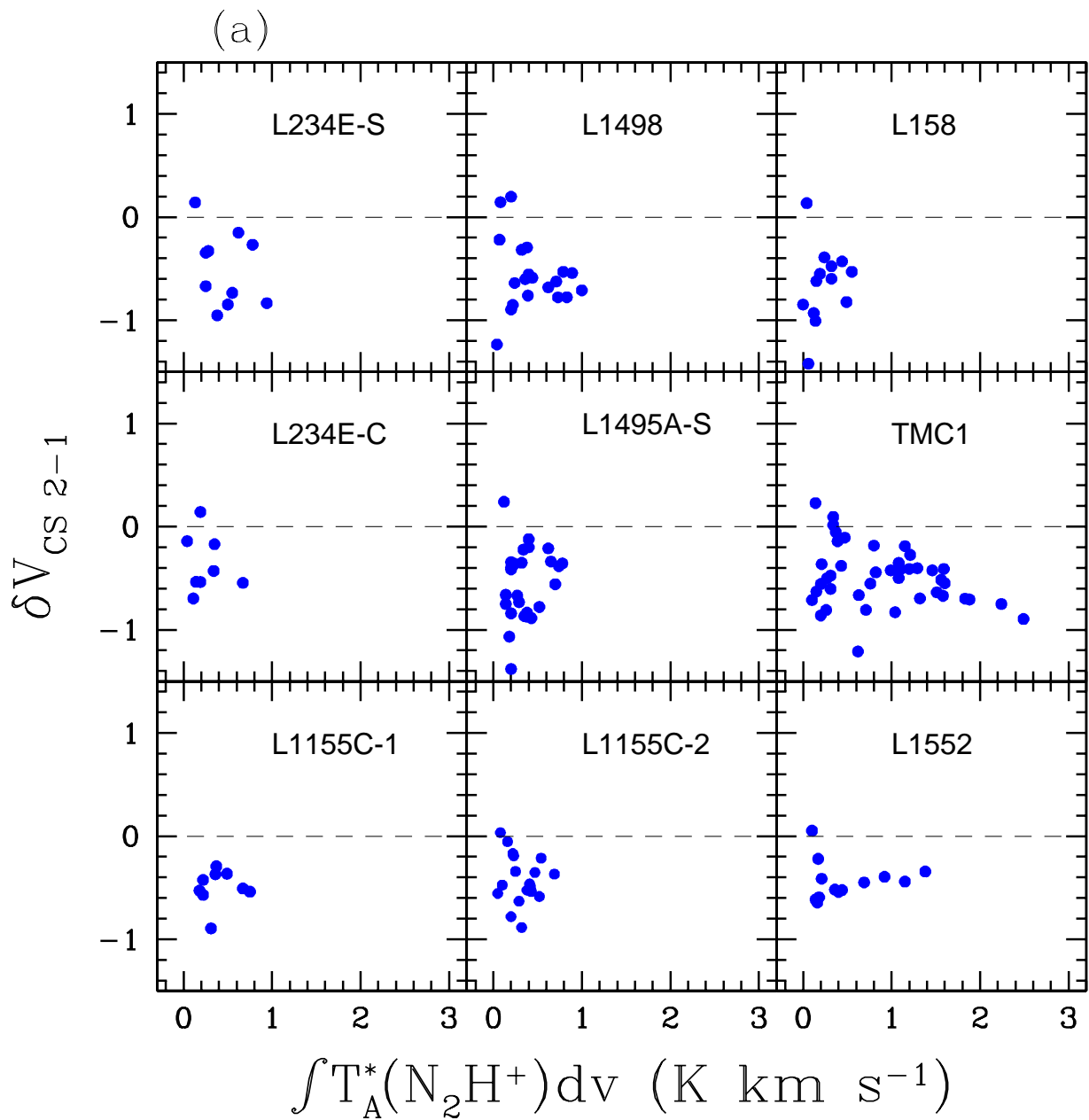}
\caption{
$\rm \delta V_{CS}$ distribution against the integrated intensity of $\rm N_2H^+$ for each starless core in four groups 
of cores in (a) contracting, (b) oscillating, (c) expanding, and (d) static motions.
In panel (a) contracting cores are put in decreasing order of $\rm -<\delta V_{CS}>$. 
}
\end{figure}

\clearpage
\begin{figure}
\centering
\includegraphics[width=7.5in]{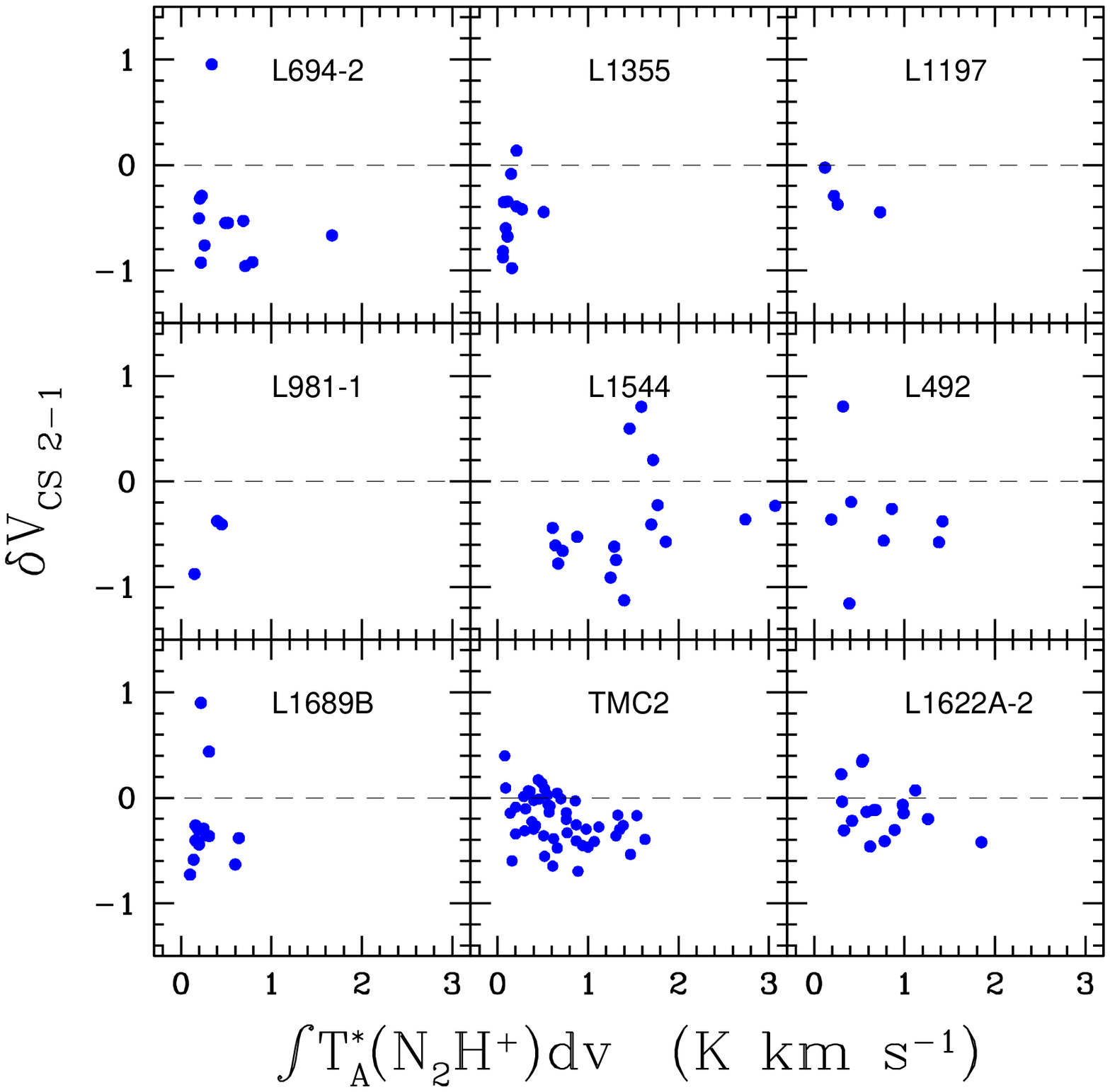}
\end{figure}

\clearpage
\begin{figure}
\centering
\includegraphics[width=7.5in]{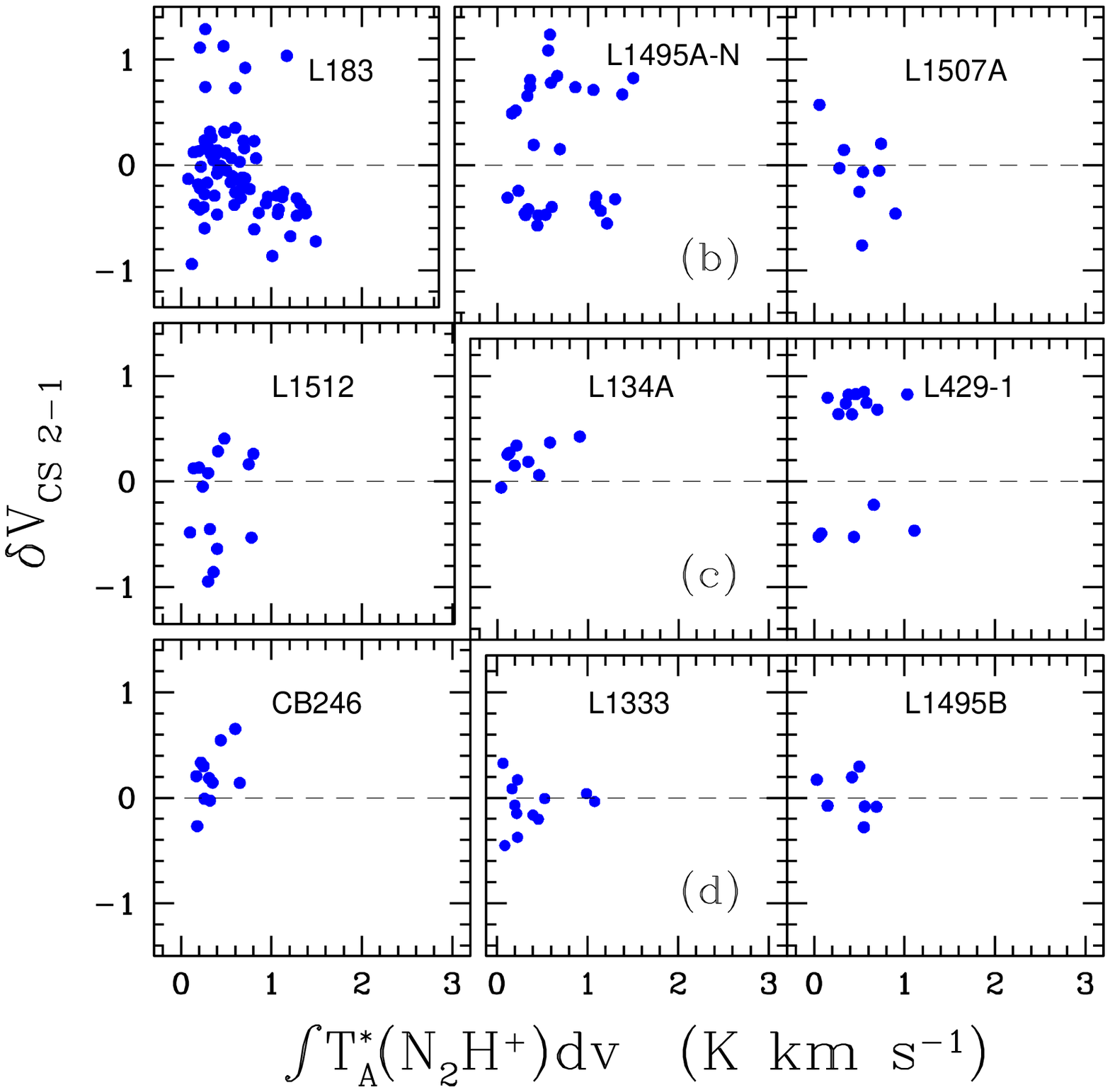}
\end{figure}

\clearpage
\begin{figure}
\centering
\includegraphics[width=7.5in]{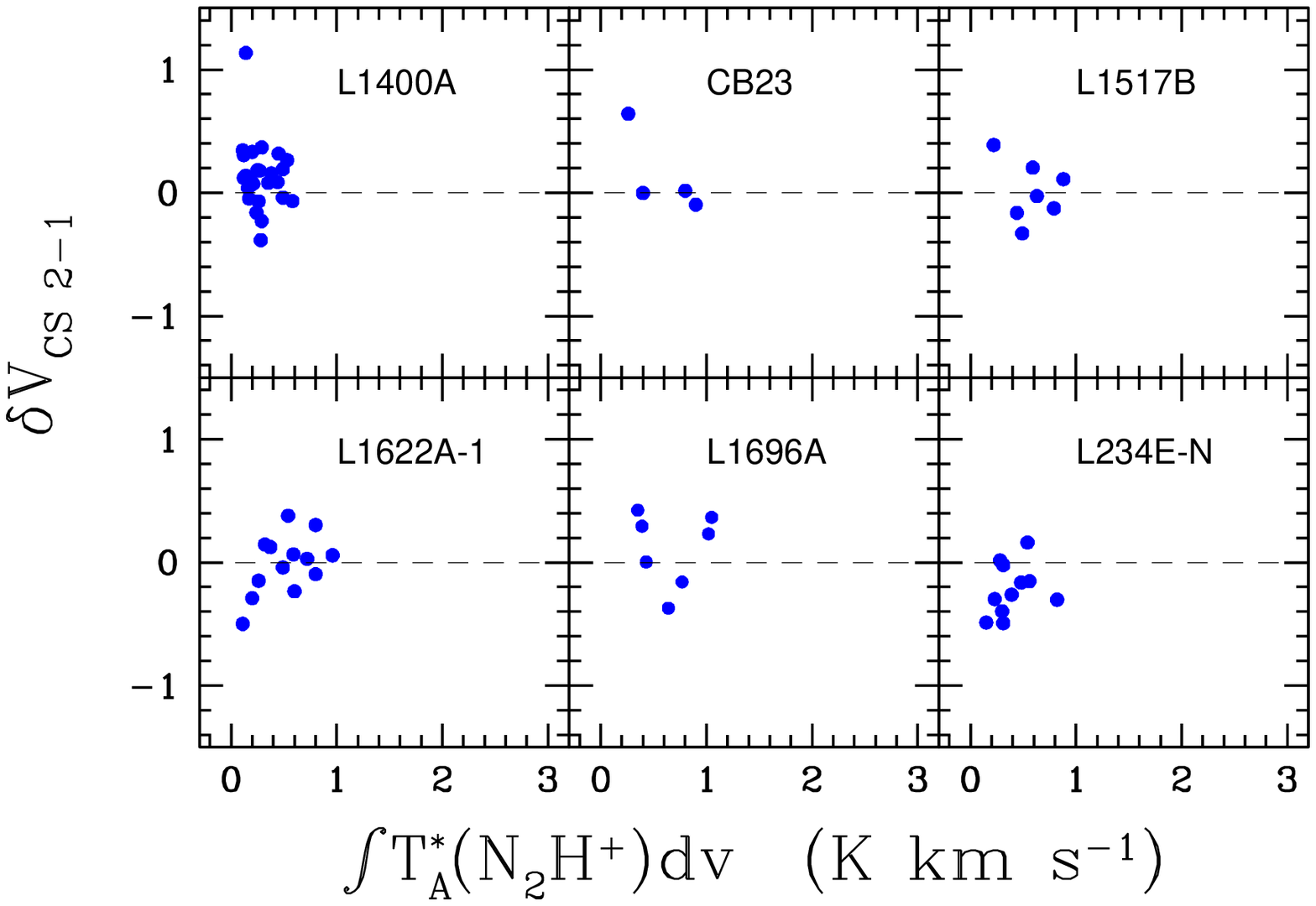}
\end{figure}

\clearpage
\begin{figure}
\centering
\includegraphics[height=7.5in,angle=0]{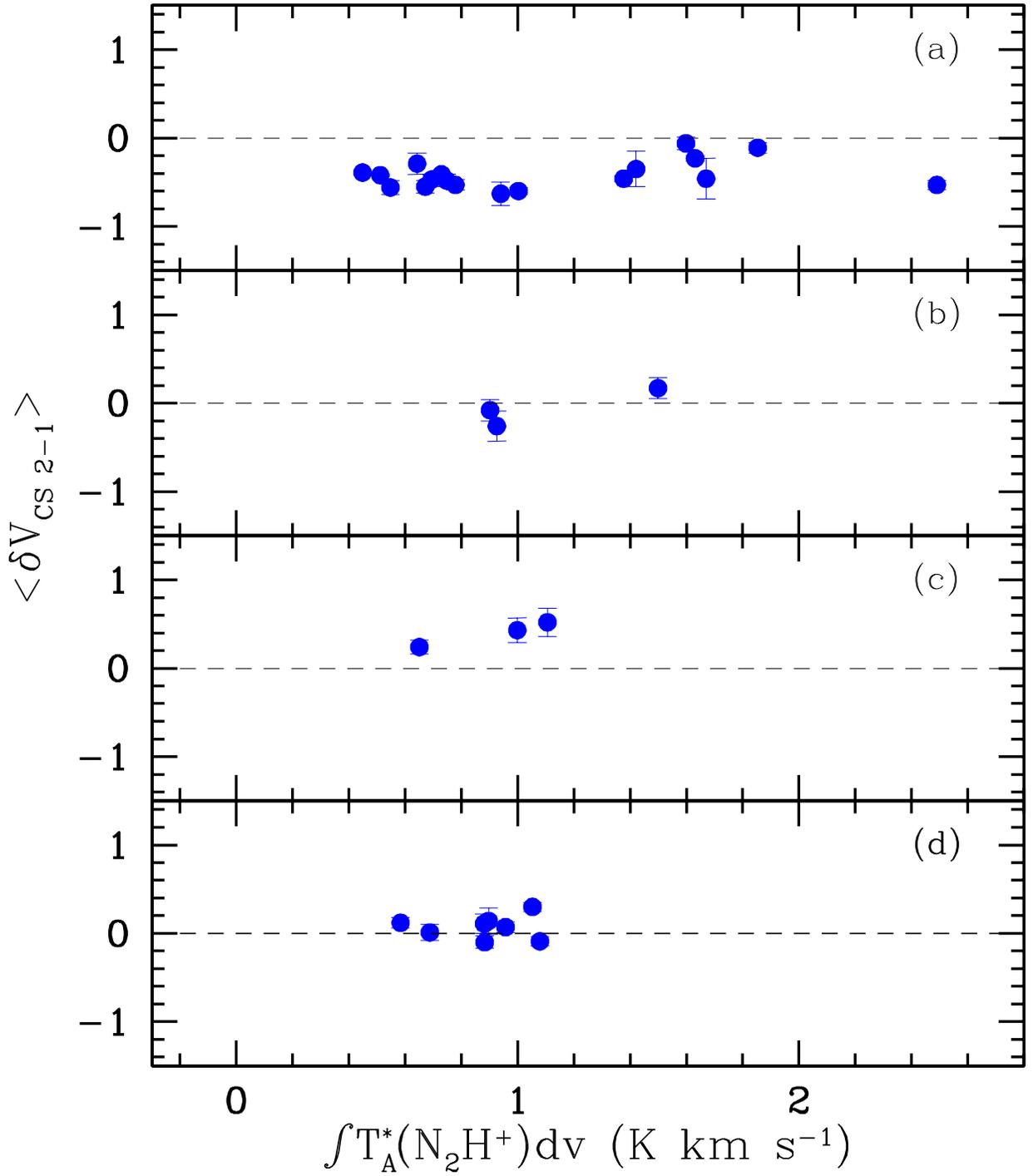}
\caption{
$\rm \delta V_{CS}$ distribution against the integrated intensity of $\rm N_2H^+$ for (a) 19 contracting cores,
(b) 3 oscillating cores, (c) 3 expanding cores, and (d) 8 static cores.
}
\end{figure}

\small
 
\def\etal{{\it et al.~}}
\def\msun{{\rm\,M_\odot}}
\def\today{\rightline{\ifcase\month\or
        January\or February\or March\or April\or May\or June\or
        July\or August\or September\or October\or November\or December\fi
        \space\number\day, \number\year}}


\begin{deluxetable}{llcccc}
\setlength{\tabcolsep}{0.9mm} 
\tablewidth{56pc}
\tablewidth{0pt}
\tablecolumns{6} 

\tablecaption{Data Summary.\tablenotemark{a}}
\tablehead{
        \colhead{Observing Line}&    
        \colhead{Telescope }     & \colhead{Observing Mode}     & 
        \colhead{Number of Cores} & \colhead{Number of Cores} & \colhead{Reference}  
        }
\startdata
$\rm N_2H^+$ 1-0     & Haystack 37m & Single pointing &   72   &   72  &  1  \nl
CS 2-1               & Haystack 37m & Single pointing &  163   &   66  &  1  \nl
CS 3-2               & NRAO 12m     & Single pointing &   91   &   66  &  2  \nl
HCN 1-0              & TRAO 14m     & Single pointing &   65   &   48  &  3  \nl
$\rm N_2H^+$ 1-0     & FCRAO 14m    & Mapping         &   35   &   35  &  4  \nl
CS 2-1               & FCRAO 14m    & Mapping         &   50   &   34  &  4  \nl
\enddata                              									       

\tablenotetext{a}{ 
Summary of the data used in this paper.
Column 1 and 2 list the observed molecular lines and the telescopes. Note that NRAO 12m is now referred to KP12m, operated by Arizona Radio Observatory.
Column 3 explains survey modes that we used, single pointing observations toward central regions of the cores and 
mapping observations fully covering the cores.
Column 4 gives the number of sources detected with each observing line in Column 1. Here in the number we dropped L1521F and L673-7 because they 
are now known to have a faint embedded source from $\it Spitzer$'s observations (Bourke et al. 2006; Dunham et al. 2010).
Column 5 provides the number of sources detected with both the line in Column 1 as optically thick tracer and $\rm N_2H^+$ as optically thin tracer.    
Here the sources detected in each set of observed lines are not exactly the same although most of sources are commonly detected in those lines.
Column 6 lists references from which the data are obtained - (1) Lee et al. 1999; (2) Lee et al. 2004; (3) Sohn et al. 2007; (4) LMT01.
}
\end{deluxetable}											       


 
\def\etal{{\it et al.~}}
\def\msun{{\rm\,M_\odot}}
\def\today{\rightline{\ifcase\month\or
        January\or February\or March\or April\or May\or June\or
        July\or August\or September\or October\or November\or December\fi
        \space\number\day, \number\year}}


\begin{deluxetable}{llrrcccccccc}
\small
\setlength{\tabcolsep}{0.9mm} 
\tablewidth{56pc}
\tablewidth{0pt}
\tablecolumns{12} 

\tablecaption{Basic Parameters of 33 Starless $\rm N_2H^+$ Cores.\tablenotemark{a}}
\tablehead{
        \colhead{Source}&    
        \colhead{R.A.}     & \colhead{Decl.}     & 
        \colhead{ $\rm <\delta V_{CS}>$} & \colhead{$\rm \int  T^*_A(N_2H^+) dv$} & 
        \colhead{ N } & 
        \colhead {$\rm N_{5\sigma}$} & \colhead{ $\rm N_p$ } & \colhead{$\rm N_n$} & 
        \colhead{$\rm N_{H_2}$} & 
        \colhead{ $\rm R_{N_2H^+}$} &
        \nl
        \colhead{}      &   
        \colhead{(2000)}     & \colhead{(2000)}     &
        \colhead{ } & \colhead{($\rm K~km~s^{-1}$)} & 
        \colhead{} & 
        \colhead{ } & \colhead{} & \colhead{ } & 
        \colhead{($\rm \times 10^{21}~cm^{-2}$)}  & 
        \colhead{ (pc) } & 
        }
\startdata
L1333    & 02~26~13.8& 75~27~02& -0.09$\pm$ 0.05 &1.08$\pm$  0.04 & 12  &  9 &  4  &  8 &   3.0 &($1^{'}.2$)\nl
L1355    & 02~53~12.2& 68~55~52& -0.42$\pm$ 0.02 &0.51$\pm$  0.04 & 12  &  3 &  1  & 11 &   1.4 &  0.05      \nl
L1498    & 04~10~51.5& 25~09~50& -0.60$\pm$ 0.04 &1.00$\pm$  0.05 & 20  & 13 &  2  & 18 &   2.8 &  0.08      \nl
L1495B   & 04~18~05.1& 28~22~22&  0.01$\pm$ 0.10 &0.69$\pm$  0.06 &  7  &  5 &  3  &  4 &   1.9 &  0.03      \nl
L1495A-N & 04~15~25.5& 28~20~15&  0.17$\pm$ 0.13 &1.50$\pm$  0.07 & 29  & 25 & 15  & 14 &   4.1 &  0.11      \nl
L1495A-S & 04~15~35.6& 28~26~35& -0.53$\pm$ 0.06 &0.78$\pm$  0.04 & 26  & 19 &  1  & 25 &   2.2 &  0.14      \nl
L1400A   & 04~30~56.8& 54~52~36&  0.12$\pm$ 0.06 &0.58$\pm$  0.09 & 25  & 10 & 18  &  7 &   1.6 &  0.10      \nl
TMC2     & 04~32~48.7& 24~24~12& -0.23$\pm$ 0.03 &1.63$\pm$  0.06 & 49  & 42 & 10  & 39 &   4.5 &  0.13      \nl
TMC1     & 04~41~33.0& 25~44~44& -0.53$\pm$ 0.05 &2.49$\pm$  0.06 & 42  & 29 &  3  & 39 &   6.9 &  0.15      \nl
L1507A   & 04~42~38.6& 29~43~45& -0.08$\pm$ 0.13 &0.90$\pm$  0.19 &  9  &  0 &  3  &  6 &   2.5 &  0.05      \nl
CB23     & 04~43~27.7& 29~39~11&  0.14$\pm$ 0.17 &0.90$\pm$  0.21 &  4  &  0 &  2  &  2 &   2.5 &\nodata     \nl
L1517B   & 04~55~18.8& 30~38~04&  0.11$\pm$ 0.10 &0.88$\pm$  0.15 &  7  &  1 &  3  &  4 &   2.4 &  0.04      \nl
L1512    & 05~04~06.2& 32~43~09& -0.26$\pm$ 0.18 &0.93$\pm$  0.07 & 14  &  9 &  7  &  7 &   2.6 &  0.07      \nl
L1544    & 05~04~14.9& 25~11~08& -0.36$\pm$ 0.06 &2.90$\pm$  0.31 & 17  &  5 &  3  & 14 &   8.0 &  0.05      \nl
L1552    & 05~17~36.0& 26~05~18& -0.46$\pm$ 0.03 &1.38$\pm$  0.05 & 13  &  7 &  1  & 12 &   3.8 &  0.06      \nl
l1622A-2 & 05~54~38.8& 01~53~44& -0.11$\pm$ 0.06 &1.85$\pm$  0.10 & 17  & 15 &  4  & 13 &   5.1 &  0.27      \nl
l1622A-1 & 05~54~53.5& 01~57~24&  0.07$\pm$ 0.06 &0.96$\pm$  0.08 & 13  &  9 &  7  &  6 &   2.7 &  0.25      \nl
L134A    & 15~53~33.1&-04~35~26&  0.43$\pm$ 0.14 &1.00$\pm$  0.22 &  9  &  1 &  8  &  1 &   2.8 &  0.08      \nl
L183     & 15~54~06.5&-02~52~23& -0.06$\pm$ 0.07 &1.60$\pm$  0.06 & 76  & 52 & 29  & 47 &   4.4 &  0.28      \nl
L1696A   & 16~28~31.4&-24~19~08&  0.30$\pm$ 0.05 &1.05$\pm$  0.15 &  7  &  2 &  5  &  2 &   2.9 &  0.04      \nl
L1689B   & 16~34~45.8&-24~37~51& -0.29$\pm$ 0.13 &0.64$\pm$  0.04 & 14  &  7 &  2  & 12 &   1.8 &  0.09      \nl
L158     & 16~47~23.2&-13~58~37& -0.56$\pm$ 0.08 &0.55$\pm$  0.07 & 13  &  4 &  1  & 12 &   1.5 &  0.07      \nl
L234E-N  & 16~48~02.0&-10~47~08& -0.10$\pm$ 0.08 &0.88$\pm$  0.08 & 11  &  5 &  2  &  9 &   2.4 &  0.08      \nl
L234E-C  & 16~48~02.0&-10~49~20& -0.55$\pm$ 0.07 &0.67$\pm$  0.08 &  8  &  1 &  1  &  7 &   1.9 &  0.04      \nl
L234E-S  & 16~48~08.6&-10~57~25& -0.63$\pm$ 0.14 &0.94$\pm$  0.07 & 10  &  6 &  1  &  9 &   2.6 &  0.06      \nl
L492     & 18~15~46.1&-03~46~13& -0.35$\pm$ 0.21 &1.42$\pm$  0.06 &  8  &  7 &  1  &  7 &   3.9 &  0.09      \nl
L429-1   & 18~17~05.6&-08~13~30&  0.45$\pm$ 0.17 &1.11$\pm$  0.05 & 16  & 11 & 11  &  5 &   3.1 &  0.11      \nl
L694-2   & 19~41~04.5& 10~57~02& -0.46$\pm$ 0.25 &1.67$\pm$  0.07 & 12  &  7 &  1  & 11 &   4.6 &  0.09      \nl
L1155C-2 & 20~42~58.8& 67~48~18& -0.47$\pm$ 0.06 &0.70$\pm$  0.04 & 18  & 14 &  1  & 17 &   1.9 &  0.31      \nl
L1155C-1 & 20~43~30.0& 67~52~42& -0.48$\pm$ 0.08 &0.75$\pm$  0.05 &  9  &  7 &  0  &  9 &   2.1 &  0.15      \nl
L981-1   & 21~00~13.2& 50~20~50& -0.39$\pm$ 0.02 &0.45$\pm$  0.04 &  3  &  2 &  0  &  3 &   1.2 &  0.09      \nl
L1197    & 22~37~02.3& 58~57~21& -0.41$\pm$ 0.04 &0.73$\pm$  0.06 &  4  &  2 &  0  &  4 &   2.0 &  0.10      \nl
CB246    & 23~56~49.2& 58~34~29&  0.24$\pm$ 0.09 &0.65$\pm$  0.04 & 11  &  8 &  8  &  3 &   1.8 &  0.10      \nl
\enddata                              									       

\tablenotetext{a}{ 
Summary of useful information of the spectral lines of 33 $\rm N_2H^+$ starless cores discussed in this paper.
Most of the data here are from tables in LMT01, but  with slight revision from more careful reduction of the spectra. 
Column 1 lists the names of $\rm N_2H^+$ cores. Columns 2 and 3 indicate the coordinates of  the positions of 
peak $\rm N_2H^+$ intensity in the cores in J2000 epoch. Column 4 gives the mean $\rm \delta V_{CS}$ of the spectra in the cores for
which S/N ratio in the integrated intensity of  $\rm N_2H^+$ is larger than 5.  
Column 5 lists the integrated intensity of the brightest $\rm N_2H^+$ position in each core. Column 6 is 
the total number of positions in the core where 
$\rm \delta V_{CS}$ of the spectra was measured. 
Column 7 lists the number of positions
for which the S/N ratio in the integrated intensity of $\rm N_2H^+$ is larger than 5.
Columns 8 and 9 are the numbers of positions in the core where 
$\rm \delta V_{CS}$ is positive and negative, respectively. 
Column 10 and 11 list column density of $\rm H_2$ at the  brightest $\rm N_2H^+$ position in each core which is 
derived as described in Section 4.1, and one-half the largest separation between positions having $\rm N_2H^+$ spectra with peak S/N 
larger than 5, respectively. The `half' separation for L1333 is given in angular scale because its distance is not well known.
}
\end{deluxetable}											       


\end{document}